# Representation Results for Defeasible Logic

G. Antoniou, D. Billington, G. Governatori and M.J. Maher

School of Computing and Information Technology, Griffith University

Nathan, QLD 4111, Australia

{ga,db,guido,mjm}@cit.gu.edu.au

**Abstract**

The importance of transformations and normal forms in logic programming, and generally in computer science, is well documented. This paper investigates transformations and normal forms in the context of Defeasible Logic, a simple but efficient formalism for nonmonotonic reasoning based on rules and priorities. The transformations described in this paper have two main benefits: on one hand they can be used as a theoretical tool that leads to a deeper understanding of the formalism, and on the other hand they have been used in the development of an efficient implementation of defeasible logic.

## 1 Introduction

Normal forms play an important role in computer science. Examples of areas where normal forms have proved fruitful include logic, where normal forms of formulae are used both for the proof of theoretical results and in automated theorem proving, and relational databases [7], where normal forms have been the driving force in the development of database theory and principles of good data modelling.

In computer science, usually normal forms are supported by *transformations*, operational procedures that transform initial objects (such as programs or logical theories) to their normal form. Such transformations are important for two main reasons:

1. They support the understanding and assimilation of new concepts because they allow one to concentrate on certain forms and key features only. Thus transformations can be useful as theoretical tools.

2. They support the optimized execution of a language; therefore they can facilitate the development of algorithms (see resolution [19]; for the significance of transformations in logic programming see [17]). Thus transformations are also useful in implementations.

In this paper we will study transformations in the setting of a particular logical formalism, *Defeasible Logic* [15, 16], following the presentation of [6]. Benefits flowing out of our results fall into both categories mentioned above.

*Defeasible Logic* is an approach to nonmonotonic reasoning [1, 14] that has a very distinctive feature: It was designed to be easily implementable right from the beginning, unlike most other approaches. Recent implementations include *Deimos* [20, 13], a query answering system capable of dealing with 100,000s of defeasible rules; and *Delores* [13], a system that calculates all conclusions, which makes use of the transformations presented in this paper.

Defeasible Logic is historically the first of a family of approaches based on the idea of *logic programming without negation as failure*. More recent logics in this family include courteous logic programs [10] and LPwNF [9]; interestingly Defeasible Logic was shown to be the most expressive of these systems (w.r.t. sceptical reasoning) [5]. This family of approaches has recently attracted considerable interest. Apart from implementability, its use in various application domains has been advocated, including the modelling of regulations and business rules [11, 3], modelling of contracts [18], and the integration of information from various sources [4].

There are five kinds of features in Defeasible Logic: facts, strict rules, defeasible rules, defeaters, and a superiority relation among rules. Essentially the superiority relation provides information about the relative strength of rules, that is, it provides information about which rules can overrule which other rules. A program or knowledge base that consists of these items is called a *defeasible theory*.

After an introduction of Defeasible Logic in section 2, in section 3 we conduct a detailed study of the proof theory of Defeasible Logic. As a consequence, we show that for every defeasible theory $T$ there is an equivalent theory $T'$ which has an empty superiority relation, and neither defeaters nor facts. However in section 3 no insight is provided as to how $T'$ might be computed by transforming the original theory $T$. This question is the driving force for the remainder of the paper.

In section 4 we introduce two key properties of transformations, modularity and incrementality. Essentially *modularity* says that a transformation may be applied to each unit of information, independent of its context; stated another way, a modular transformation may be applied to a part of a program or theory without the need to notify or modify the rest. And *incrementality* says that if a theory has been transformed, then an update in the original theory should have cost proportional to the change, without the need to transform the entire updated theory anew. Obviously both properties are important for implementations.

After establishing these properties we proceed to present in section 5, the main section of this paper, transformations that

- normalize a theory by eliminating facts and separating the definite and defeasible reasoning levels as much as possible;

- eliminate defeaters;

- lead to an empty superiority relation,

without changing the meaning of the defeasible theory (in the original language). For each of these transformations we study which of the two key properties they satisfy. Moreover, in case they do not satisfy modularity or incrementality, we show that there is no other (correct) transformation that satisfies the property. Finally we present a significantly simplified proof theory that can be used once all these transformations have been applied.

# 2 Basics of Defeasible Logic

## 2.1 An Informal Presentation

We begin by presenting the basic ingredients of Defeasible Logic. A defeasible theory (a knowledge base in Defeasible Logic, or a defeasible logic program) consists of five different kinds of knowledge: facts, strict rules, defeasible rules, defeaters, and a superiority relation.

*Facts* are indisputable statements, for example, "Tweety is an emu". Written formally, this would be expressed as $emu(tweety)$.

*Strict rules* are rules in the classical sense: whenever the premises are indisputable (e.g. facts) then so is the conclusion. An example of a strict rule is "Emus are birds". Written formally:

$$emu(X) \rightarrow bird(X).$$

*Defeasible rules* are rules that can be defeated by contrary evidence. An example of such a rule is "Birds typically fly"; written formally:

$$bird(X) \Rightarrow flies(X).$$

The idea is that if we know that something is a bird, then we may conclude that it flies, *unless there is other, not inferior, evidence suggesting that it may not fly.*

*Defeaters* are rules that cannot be used to draw any conclusions. Their only use is to prevent some conclusions. In other words, they are used to defeat some defeasible rules by producing evidence to the contrary. An example is "If an animal is heavy then it might not be able to fly". Formally:

$$heavy(X) \leadsto \neg flies(X).$$

The main point is that the information that an animal is heavy is not sufficient evidence to conclude that it doesn't fly. It is only evidence that the animal *may* not be able to fly. In other words, we don't wish to conclude $\neg flies(X)$ if $heavy(X)$, we simply want to prevent a conclusion $flies(X)$.

The *superiority relation* among rules is used to define priorities among rules, that is, where one rule may override the conclusion of another rule. For example, given the defeasible rules

$$\begin{aligned} r: & \quad bird(X) \Rightarrow flies(X) \\ r': & \quad brokenWing(X) \Rightarrow \neg flies(X) \end{aligned}$$

which contradict one another, no conclusive decision can be made about whether a bird with broken wings can fly. But if we introduce a superiority relation $>$ with $r' > r$, then we can indeed conclude that the bird cannot fly.

Notice that a cycle in the superiority relation is counter-intuitive. In the above example, it makes no sense to have both $r > r'$ and $r' > r$. Consequently, we will focus on cases where the superiority relation is acyclic.

Another point worth noting is that, in Defeasible Logic, priorities are *local* in the following sense: Two rules are considered to be competing with one another only if they have complementary heads. Thus, since the superiority relation is used to resolve conflicts among competing rules, it is only used to compare rules with complementary heads; the information $r > r'$ for rules $r, r'$ without complementary heads may be part of the superiority relation, but has no effect on the proof theory.

## 2.2 Formal Definition

In this paper we restrict attention to essentially propositional Defeasible Logic. Rules with free variables are interpreted as rule schemas, that is, as the set of all ground instances; in such cases we assume that the Herbrand universe is finite. We assume that the reader is familiar with the notation and basic notions of propositional logic. If $q$ is a literal, $\sim q$ denotes the complementary literal (if $q$ is a positive literal $p$ then $\sim q$ is $\neg p$; and if $q$ is $\neg p$, then $\sim q$ is $p$).

Rules are defined over a *language* (or *signature*) $\Sigma$, the set of propositions (atoms) and labels that may be used in the rule.

A *rule* $r : A(r) \hookrightarrow C(r)$ consists of its unique *label* $r$, its *antecedent* $A(r)$ ($A(r)$ may be omitted if it is the empty set) which is a finite set of literals, an arrow, and its *head* (or *consequent*) $C(r)$ which is a literal. In writing rules we omit set notation for antecedents and sometimes we omit the label when it is not relevant for the context. There are three kinds of rules, each represented by a different arrow. Strict rules use $\rightarrow$, defeasible rules use $\Rightarrow$, and defeaters use $\rightsquigarrow$.

Given a set $R$ of rules, we denote the set of all strict rules in $R$ by $R_s$, the set of strict and defeasible rules in $R$ by $R_{sd}$, the set of defeasible rules in $R$ by $R_d$, and the set of defeaters in $R$ by $R_{dft}$. $R[q]$ denotes the set of rules in $R$ with consequent $q$.

A *superiority relation on* $R$ is a relation $>$ on $R$. When $r_1 > r_2$, then $r_1$ is called *superior* to $r_2$, and $r_2$ *inferior* to $r_1$. Intuitively, $r_1 > r_2$ expresses that $r_1$ overrules $r_2$, should both rules be applicable. Typically we assume $>$ to be acyclic (that is, the transitive closure of $>$ is irreflexive), but in this paper we occasionally study which properties depend on acyclicity.

A *defeasible theory* $D$ is a triple $(F, R, >)$ where $F$ is a finite set of literals (called *facts*), $R$ a finite set of rules, and $>$ a superiority relation on $R$. $D$ is called *well-formed* iff $>$ is acyclic and $>$ is only defined on rules with complementary heads. $D$ is called *cyclic* iff $>$ is cyclic. In case $F = \emptyset$ and $>= \emptyset$, we denote a defeasible theory $(\emptyset, R, \emptyset)$ by $R$.

The *language* (or *signature*) *of* $D$ is the set of propositions and labels $\Sigma$ that are used within $D$. In cases where it is unimportant to refer to the language of $D$, $\Sigma$ will not be mentioned.

## 2.3 Proof Theory

A *conclusion* of $D$ is a tagged literal and can have one of the following four forms:

$+\Delta q$ which is intended to mean that $q$ is definitely provable in $D$.

$-\Delta q$ which is intended to mean that we have proved that $q$ is not definitely provable in $D$.

$+\partial q$ which is intended to mean that $q$ is defeasibly provable in $D$.

$-\partial q$ which is intended to mean that we have proved that $q$ is not defeasibly provable in $D$.

In section 3 we will discuss the interconnections of these concepts. At this stage we wish to mention only one: If we are able to prove $q$ definitely, then $q$ is also defeasibly provable. This is a direct consequence of the formal definition below. It resembles the situation in, say, default logic: a formula is sceptically provable from a default theory $T = (W, D)$ (in the sense that it is included in each extension) if it is provable from the set of facts $W$.

Provability is defined below. It is based on the concept of a *derivation* (or *proof*) in $D = (F, R, >)$. A derivation is a finite sequence $P = (P(1), \ldots P(n))$ of tagged literals satisfying the following conditions ($P(1..i)$ denotes the initial part of the sequence $P$ of length $i$):

$+\Delta$: If $P(i+1) = +\Delta q$ then either
    $q \in F$ or
    $\exists r \in R_s[q] \; \forall a \in A(r) : +\Delta a \in P(1..i)$

That means, to prove $+\Delta q$ we need to establish a proof for $q$ using facts and strict rules only. This is a deduction in the classical sense – no proofs for the negation of $q$ need to be considered (in contrast to defeasible provability below, where opposing chains of reasoning must be taken into account, too).

$-\Delta$: If $P(i+1) = -\Delta q$ then
    $q \notin F$ and
    $\forall r \in R_s[q] \; \exists a \in A(r) : -\Delta a \in P(1..i)$

To prove $-\Delta q$, i.e. that $q$ is not definitely provable, $q$ must not be a fact. In addition, we need to establish that every strict rule with head $q$ is *known to be* inapplicable. Thus for every such rule $r$ there must be at least one antecedent $a$ for which we have established that $a$ is not definitely provable $(-\Delta a)$.

It is worth noticing that this definition of nonprovability does not involve loop detection. Thus if $D$ consists of the single rule $p \to p$, we can see that $p$ cannot be proven, but Defeasible Logic is unable to prove $-\Delta p$.

$+\partial$: If $P(i+1) = +\partial q$ then either
    (1) $+\Delta q \in P(1..i)$ or
    (2) (2.1) $\exists r \in R_{sd}[q] \forall a \in A(r) : +\partial a \in P(1..i)$ and
        (2.2) $-\Delta \sim q \in P(1..i)$ and
        (2.3) $\forall s \in R[\sim q]$ either
            (2.3.1) $\exists a \in A(s) : -\partial a \in P(1..i)$ or
            (2.3.2) $\exists t \in R_{sd}[q]$ such that
                $\forall a \in A(t) : +\partial a \in P(1..i)$ and $t > s$

Let us illustrate this definition. To show that $q$ is provable defeasibly we have two choices: (1) We show that $q$ is already definitely provable; or (2) we need to argue using the defeasible part of $D$ as well. In particular, we require that there must be a strict or defeasible rule with head $q$ which can be applied (2.1). But now we need to consider possible "attacks", that is, reasoning chains in support of $\sim q$. To be more specific: to prove $q$ defeasibly we must show that $\sim q$ is not definitely provable (2.2). Also (2.3) we must consider the set of all rules which are not known to be inapplicable and which have head $\sim q$ (note that here we consider defeaters, too, whereas they could not be used to support the conclusion $q$; this is in line with the motivation of defeaters given in subsection 2.1). Essentially each such rule $s$ attacks the conclusion $q$. For $q$ to be provable, each such rule $s$ must be counterattacked by a rule $t$ with head $q$ with the following properties: (i) $t$ must be applicable at this point, and (ii) $t$ must be stronger than $s$. Thus each attack on the conclusion $q$ must be counterattacked by a stronger rule.

The definition of the proof theory of Defeasible Logic is completed by the condition $-\partial$. It is nothing more than a strong negation of the condition $+\partial$.

$-\partial$: If $P(i+1) = -\partial q$ then
    (1) $-\Delta q \in P(1..i)$ and
    (2) (2.1) $\forall r \in R_{sd}[q] \; \exists a \in A(r) : -\partial a \in P(1..i)$ or
        (2.2) $+\Delta \sim q \in P(1..i)$ or
        (2.3) $\exists s \in R[\sim q]$ such that
            (2.3.1) $\forall a \in A(s) : +\partial a \in P(1..i)$ and
            (2.3.2) $\forall t \in R_{sd}[q]$ either
                $\exists a \in A(t) : -\partial a \in P(1..i)$ or $t \not> s$

To prove that $q$ is not defeasibly provable, we must first establish that it is not definitely provable. Then we must establish that it cannot be proven using the defeasible part of the theory. There are three possibilities to achieve this: either we have established that none of the (strict and defeasible) rules with head $q$ can be applied (2.1); or $\sim q$ is definitely provable (2.2); or there must be an applicable rule $s$ with head $\sim q$ such that no possibly applicable rule $t$ with head $q$ is superior to $s$ (2.3).

The elements of a derivation are called *lines* of the derivation. We say that a tagged literal $L$ is *provable* in $D = (F, R, >)$, denoted $D \vdash L$, iff there is a derivation in $D$ such that $L$ is a line of $P$. When $D$ is obvious from the context we write $\vdash L$.

It is instructive to consider the conditions $+\partial$ and $-\partial$ in the terminology of *teams*, borrowed from [10]. At some stage there is a team $A$ consisting of the applicable rules with head $q$, and a team $B$ consisting of the applicable rules with head $\sim q$. These teams compete with one another. Team $A$ wins iff every rule in team $B$ is overruled by a rule in team $A$; in that case we can prove $+\partial q$. Another case is that team $B$ wins, in which case we can prove $+\partial \sim q$. But there are several intermediate cases, for example one in which we can prove that neither $q$ nor $\sim q$ are provable. And there are cases where nothing can be proved (due to loops). A thorough discussion of the possible outcomes of this "battle" between the two competing teams is found in the next section.

**Example 1** Here we wish to give an example[1] which illustrates the notion of teams.

$monotreme(platypus)$

$hasFur(platypus)$

$laysEggs(platypus)$

$hasBill(platypus)$

$r_1: \ monotreme(X) \Rightarrow mammal(X)$

$r_2: \ hasFur(X) \Rightarrow mammal(X)$

$r_3: \ laysEggs(X) \Rightarrow \neg mammal(X)$

$r_4: \ hasBill(X) \Rightarrow \neg mammal(X)$

$r_1 > r_3$

$r_2 > r_4$

Intuitively we conclude that *platypus* is a *mammal* because for every reason against this conclusion ($r_3$ and $r_4$) there is a stronger reason for *mammal(platypus)* ($r_1$ and $r_2$ respectively). It is easy to see that $+\partial mammal(platypus)$ is indeed provable in Defeasible Logic: there is a rule in support of *mammal(platypus)*, and every rule for $\neg mammal(platypus)$ is overridden by a rule for *mammal(platypus)*.

We conclude this section with two remarks. First, strict rules are used in two different ways. When we try to establish *definite provability*, then strict rules are used as in classical logic: if they can fire they are applied, regardless of any reasoning chains with the opposite conclusion. But strict rules can also be used to show *defeasible provability*, given that some other literals are known to be defeasible provable. In this case, strict rules are used exactly like defeasible rules. For example, a strict rule may be applicable yet it may not fire because there is a stronger rule with the opposite conclusion. Also, strict rules are not automatically superior to defeasible rules. This treatment of strict rules may look a bit confusing and counterintuitive. In subsection 5.1 we establish a simple normal form which separates the strict and the defeasible part as much as possible, with only a transparent "bridge" being allowed linking the two parts together.

---

[1]Rules in this example are actually rule schemas. Since there are no function symbols and only a finite number of propositional constants, it is still essentially a propositional example.

Finally, in the above definition often we refer to $P(1..i)$, or, intuitively, to the fact that a rule is *currently* applicable. This may create the wrong impression that this applicability may change as the proof proceeds (something found often in nonmonotonic proofs). But the sceptical nature of Defeasible Logic does not allow for such a situation. For example, if we have established that a rule is currently not applicable because we have $-\partial a$ for some antecedent $a$, this means that we have proven at a previous stage that *a is not provable from the defeasible theory D per se.*

## 3 An Analysis of the Proof Theory

We have seen in the previous section that for every proposition $p$ we have the concepts $\vdash +\Delta p, \vdash -\Delta p, \vdash +\partial p, \vdash -\partial p$ and the complementary concepts $\nvdash +\Delta p$ etc. Finally, there are the corresponding concepts for $\neg p$, which we would expect to be related to those for $p$. This section sheds light on the interrelations between these notions.

We define four sets of literals encapsulating all the conclusions of a theory $D$.

$$+\Delta = \{p \mid D \vdash +\Delta p\}$$
$$-\Delta = \{p \mid D \vdash -\Delta p\}$$
$$+\partial = \{p \mid D \vdash +\partial p\}$$
$$-\partial = \{p \mid D \vdash -\partial p\}$$

Thus, the proof-theoretic effects of a theory are summarized in the 4-tuple $(+\Delta, -\Delta, +\partial, -\partial)$.

We define two defeasible theories $D_1$ and $D_2$ to be *conclusion equivalent* if the two theories produce identical 4-tuples. We write $D_1 \equiv D_2$.

As straightforward consequences of the proof rules, we have the following relations among the sets.

$$+\Delta \subseteq +\partial \qquad +\Delta \cap -\Delta = \emptyset$$
$$-\partial \subseteq -\Delta \qquad +\partial \cap -\partial = \emptyset$$
$$+\Delta \cap -\partial = \emptyset$$

The four sets might generate $2^4 = 16$ possible outcomes for a single proposition $p$. However, because of the above relations, for each proposition $p$ we can identify exactly six different possible outcomes of the proof theory. With each outcome we present a simple theory that achieves this outcome.

A: $\nvdash -\Delta p$ and $\nvdash +\partial p$
   $p \to p$

B: $\vdash +\partial p$ and $\nvdash +\Delta p$ and $\nvdash -\Delta p$
   $\Rightarrow p; p \to p$

C: $\vdash +\Delta p$ (and also $\vdash +\partial p$)
   $\to p$

D: $\vdash +\partial p$ and $\vdash -\Delta p$
$\Rightarrow p$

E: $\vdash -\Delta p$ and $\nvdash +\partial p$ and $\nvdash -\partial p$
$p \Rightarrow p$

F: $\vdash -\partial p$ (and also $\vdash -\Delta p$)
$\emptyset$, the empty theory

We can represent the outcomes in terms of a Venn diagram in Figure 1.

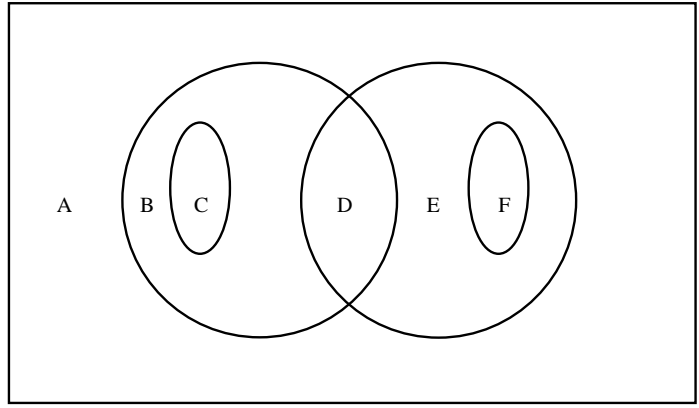

Figure 1: Proof Theory Outcomes

In Figure 1, the circle on the left – containing B, C, and D – represents the literals $p$ such that $+\partial p$ can be proved, and the ellipse inside it (i.e. C) represents the literals $p$ such that $+\Delta p$ can be proved. The circle on the right – containing D, E, and F – represents the literals $p$ such that $-\Delta p$ can be proved, and the ellipse inside it (i.e. F) represents the literals $p$ such that $-\partial p$ can be proved.

Similarly, there are the same six possibilities for $\neg p$. Due to the relationship between $p$ and $\neg p$, many fewer than the $6 \times 6 = 36$ possible combinations are possible outcomes of the proof theory. We first establish some simple results that will eliminate many combinations.

**Proposition 1** *Consider a defeasible theory D.*

1. *If $\nvdash -\Delta \neg p$ and $\nvdash +\Delta p$ then $\nvdash +\partial p$*

2. *If $\vdash +\Delta \neg p$ and $\vdash -\Delta p$ then $\vdash -\partial p$*

3. *If $\vdash +\partial \neg p$ and $\vdash -\Delta p$ and $\nvdash -\partial p$, then D is cyclic*

**Proof.** Statements 1 and 2 follow directly from the proof rules for $+\partial$ and $-\partial$. Statement 3 is proved as follows.

Suppose $\vdash +\partial \neg p$, $\vdash -\Delta p$ and $\nvdash -\partial p$. If we could prove $+\partial \neg p$ via $+\Delta \neg p$ then we could prove $-\partial p$ through (2.2) of $-\partial$, a contradiction. So there is a rule $r$ for $\neg p$ such that $\forall a \in A(r) \; +\partial a$ can be proved.

Since $\nvdash -\partial p$, by (2.3) of $-\partial$ there is a rule $t$ for $p$ such that $\forall a \in A(t) \; -\partial a$ cannot be proved, and $t > r$.

Following (2.3) of $+\partial$, there is a rule $t_1$ for $\neg p$ such that $\forall a \in A(t_1)$, $+\partial a$ can be proved and $t_1 > t$. Reconsidering (2.3) of $-\partial$, there is a rule $t_2$ for $p$ such that $\forall a \in A(t_2)$ $-\partial a$ cannot be proved, and $t_2 > t_1$.

Repeating this argument enough times, it is clear that there must be a cycle in the superiority relation, since $D$ has a finite number of rules, by definition. □

In terms of the the diagram (Figure 1), the properties of the previous proposition have the following effects (where $p$ is a positive or negative literal):

1. If $p$ satisfies A, B, D, E, or F, and $\neg p$ satisfies A, B, or C then $p$ satisfies A, E, or F (Property 1). Consequently, it is not possible for $p$ to satisfy B or D, and $\neg p$ to satisfy A, B, or C.

2. If $p$ satisfies D, E, or F, and $\neg p$ satisfies C then $p$ satisfies F (Property 2). Consequently, it is not possible for $p$ to satisfy D or E, and $\neg p$ to satisfy C.

3. If $\neg p$ satisfies B, C or D, and $p$ satisfies D or E, then $D$ is cyclic (Property 3). Consequently, if $D$ is acyclic, it is not possible for $\neg p$ to satisfy B, C or D, and $p$ to satisfy D or E.

In the following table we display the possible combinations of conclusions for a proposition $p$ and its negation $\neg p$. The table is symmetric across the leading diagonal, since the treatment of literals in Defeasible Logic – and, in particular, in the effects above – is not affected by the polarity of the literal. Those combinations which are possible are displayed as *Poss*. Those combinations which are not possible are displayed as $NP(i)$, where $i$ is the property number which implies that they are impossible. The combinations displayed as $NP(3)$ are impossible only in acyclic theories, and can be obtained for cyclic theories, as we will see shortly.

|   |   | $\neg p$ |   |   |   |   |   |
|---|---|---|---|---|---|---|---|
|   |   | A | B | C | D | E | F |
|   | A | Poss | NP(1) | Poss | NP(1) | Poss | Poss |
|   | B | NP(1) | NP(1) | NP(1) | NP(1) | NP(3) | Poss |
| $p$ | C | Poss | NP(1) | Poss | NP(1) | NP(2) | Poss |
|   | D | NP(1) | NP(1) | NP(1) | NP(3) | NP(3) | Poss |
|   | E | Poss | NP(3) | NP(2) | NP(3) | Poss | Poss |
|   | F | Poss | Poss | Poss | Poss | Poss | Poss |

It is easy to see that for all combinations that are possible, a sample theory can be obtained by combining the appropriate theories for each letter, as listed earlier.

There are five combinations that cannot be obtained by acyclic theories, but are possible when cyclic theories are permitted. These combinations are DD, BE, DE and their reverses, for which property 3 is cited in the table above. We give below an example theory for each case.

For DD

$$r_1 : \Rightarrow p$$
$$r_2 : \Rightarrow \neg p$$
$$r_1 > r_2, r_2 > r_1$$

For BE

$$\begin{aligned} r_1 : &\quad p \Rightarrow p \\ r_2 : &\quad \Rightarrow \neg p \\ r_3 : &\quad \neg p \rightarrow \neg p \\ r_1 > r_2, r_2 > r_1 \end{aligned}$$

For DE

$$\begin{aligned} r_1 : &\quad p \Rightarrow p \\ r_2 : &\quad \Rightarrow \neg p \\ r_1 > r_2, r_2 > r_1 \end{aligned}$$

From the results summarised in the above table, and the comment immediately following it, we can draw the following results.

**Theorem 2** *If $D$ is an acyclic defeasible theory, then $D$ is conclusion equivalent to a theory $D'$ that contains no use of the superiority relation, nor defeaters.*

*If $D$ is a cyclic defeasible theory, then $D$ is conclusion equivalent to a theory $D'$ that contains no use of defeaters, and if $D'$ contains cycles then they have length 2, and each cycle involves the only two rules for a literal and its complement.*

Applying the same techniques as above, we have a simple proof that the defeasible part of an acyclic defeasible theory is consistent. By consistent we mean that a theory cannot conclude that both a proposition $p$ and its negation are defeasibly true unless they are both definitely true. This was first proved by Billington in [6].

**Proposition 3** *Let $D$ be an acyclic defeasible theory. If $\vdash +\partial p$ and $\vdash +\partial \neg p$ then $\vdash +\Delta p$ and $\vdash +\Delta \neg p$. Consequently, if $D$ contains no strict rules and no facts and $\vdash +\partial q$, then $\vdash -\partial \neg q$.*

**Proof.** If $\vdash +\partial p$ and $\vdash +\partial \neg p$ then $p$ and $\neg p$ each satisfies B, C or D. Looking at the table, and since $D$ is acyclic, both $p$ and $\neg p$ satisfy C. That is, $\vdash +\Delta p$ and $\vdash +\Delta \neg p$. □

The theory above for combination DD shows that cyclic theories can be inconsistent. Furthermore, the formally greater expressiveness, in terms of combinations, of permitting cyclic superiority relations appears, from looking at the above examples, not to translate into greater usefulness. This suggests that the restriction to acyclic defeasible theories, already justified by intuition and avoiding inconsistency, provides no practical limitation.

The results of Theorem 2 are constructive in a sense, but they are not useful in implementing Defeasible Logic since they suggest a complete evaluation of the defeasible theory before the construction of an equivalent theory. In the remainder of this paper we will investigate ways of transforming an input defeasible theory to one without facts and defeaters, and with an empty superiority relation, without performing an evaluation.

## 4 Properties of Transformations

Theory transformations are an important way to exploit results of the previous section. They can be used to extend an implementation of a subset of defeasible theory to the entire theory.

A *transformation* is a mapping from defeasible theories to defeasible theories. Recall that $D_1 \equiv D_2$ iff $D_1$ and $D_2$ have the same consequences; similarly, $D_1 \equiv_\Sigma D_2$ means that $D_1$ and $D_2$ have the same consequence in the language $\Sigma$. A transformation is correct if the transformed theory has the same meaning as the original theory. Formally: a transformation $T$ is *correct* iff, for all defeasible theories $D$, $D \equiv_\Sigma T(D)$, where $\Sigma$ is the language of $D$.

Most operations on theories are minor changes to an existing theory. If the original theory has been transformed, say for efficiency reasons, then we would wish a minor update to the original theory not require that the entire updated theory be transformed. This is a form of incrementality. A transformation is incremental if the application of the transformation can be performed on a bit-by-bit basis. Formally: a transformation $T$ is *incremental* iff, for all defeasible theories $D_1$ and $D_2$, $T(D_1 \cup D_2) \equiv_\Sigma T(D_1) \cup T(D_2)$, where $\Sigma$ is the union of the languages of $D_1$ and $D_2$.

Similarly, when we change the representation of knowledge in a part of a defeasible theory, we would like this change to be invisible to the remainder of the theory. This concept of modularity is important in all forms of software development. A transformation is modular if it can be applied to a part of a theory without modifying the meaning of the theory as a whole. Formally: a transformation $T$ is *modular* iff, for all defeasible theories $D_1$ and $D_2$, $D_1 \cup D_2 \equiv_\Sigma T(D_1) \cup D_2$, where $\Sigma$ is the union of the languages of $D_1$ and $D_2$.

**Proposition 4** *If a transformation is modular then it is correct and incremental*

**Proof.** Taking $D_2 = \emptyset$ in the definition of modularity, we have $D_1 \equiv_\Sigma T(D_1)$ which expresses correctness. By modularity $T(D_1) \cup T(D_2) \equiv_\Sigma D_1 \cup T(D_2)$, again by modularity $D_1 \cup T(D_2) \equiv_\Sigma D_1 \cup D_2$, then by correctness $D_1 \cup D_2 \equiv_\Sigma T(D_1 \cup D_2)$; therefore $T(D_1) \cup T(D_2) \equiv_\Sigma T(D_1 \cup D_2)$ which expresses incrementality. □

In general the inverse of Proposition 4 does not hold. As we shall see in section 5 there are correct and incremental transformations that are not modular.

## 5 Transformations of Defeasible Theories

Previously (Theorem 2) we showed that any acyclic defeasible theory is equivalent to one which uses no defeaters and an empty superiority relation. Here we shall provide two transformations that together remove defeaters and empty the superiority relation. Both are based on the same approach. We introduce new literals that are intermediate between rule bodies and rule heads. The effects of the simulated feature in limiting inference are simulated by new defeasible rules attacking the inference of these intermediate literals. Since these literals are not in the language of the original program, any inferences made by the new rules will not affect correctness.

We begin with a normalization process that eliminates facts and makes the dual use of strict rules (in the definite and defeasible part) transparent.

### 5.1 A Normal Form for Defeasible Logic

We propose a normal form for defeasible theories. The main purpose of this normal form is to provide a separation of concerns, within a defeasible theory, between definite and defeasible conclusions. In Defeasible Logic, a strict rule may participate in the superiority relation. This participation has no effect on the definite conclusions of the theory, but can affect the defeasible conclusions. We consider theories where this occurs to be somewhat misleading, and propose a normal form in which definite and defeasible reasoning are separated as much as is practicable.

**Definition 5** *A defeasible theory $D = (F, R, >)$ is normalized (or in normal form) iff the following three conditions are satisfied:*

(a) *Every literal is defined either solely by strict rules, or by one strict rule and other non-strict rules.*

(b) *No strict rule participates in the superiority relation $>$.*

(c) *$F = \emptyset$*

Every defeasible theory can be transformed into normal form. This establishes that facts are not needed in the formulation of defeasible logic, and that the misleading theories we discussed above are unnecessary. We now define this transformation explicitly. Following that we prove that the transformation preserves the conclusions in the language of $D$.

**Definition 6** *Consider a defeasible theory $D = (F, R, >)$, and $\Sigma$ be the language of $D$. We define $normal(D) = (\emptyset, R', >)$, where $R'$ is defined below.*

*Let $'$ be a function which maps propositions to new (previously unused) propositions, and rule names to new rule names. We extend this, in the obvious way, to literals and conjunctions of literals.*

$R' = R_d \cup R_{dft} \cup$
$\quad \{\to f' \mid f \in F\} \cup$
$\quad \{r' : A' \to C' \mid r : A \to C \text{ is a strict rule in } R\} \cup$
$\quad \{r : A \Rightarrow C \mid r : A \to C \text{ is a strict rule in } R\} \cup$
$\quad \{p' \to p \mid A \to p \in R \text{ or } p \in F\}$

*The rules derived from $F$ and rules $p' \to p$ are given distinct new names.*

It is clear from the transformation described above that $normal(D)$ is normalized (i.e. satisfies conditions (a)–(c)). Notice that strict rules have been altered to become defeasible rules, although their names are unchanged. Thus although $>$ is unchanged, it now no longer concerns any strict rule.

**Example 2** Consider the following defeasible theory.

$$
\begin{aligned}
&e\\
&a\\
&r1: \quad a \to b\\
&r2: \quad \Rightarrow c\\
&r3: \quad c \to d\\
&r4: \quad e \Rightarrow \neg d\\
&r4 > r3
\end{aligned}
$$

The transformed theory looks as follows:

$$
\begin{array}{llll}
r1': & a' \to b' & r1: & a \Rightarrow b\\
r2: & \Rightarrow c & r3': & c' \to d'\\
r3: & c \Rightarrow d & r4: & e \Rightarrow \neg d\\
r5: & \to e' & r6: & \to a'\\
r7: & a' \to a & r8: & b' \to b\\
r9: & d' \to d & r10: & e' \to e\\
& & r4 > r3
\end{array}
$$

**Theorem 7** *The transformation normal is correct.*

**Proof.** We split the proof in two cases. Let $q$ be a tagged literal in the language $\Sigma$. We first prove that if $D \vdash q$ then $normal(D) \vdash q$, and then the other direction, namely: if $normal(D) \vdash q$, then $D \vdash q$.

*Case $\Rightarrow$.* We prove the property by induction on the length of proofs in $D$.

*Inductive base: $n = 1$.* In this case a proof $P$ consists of the single line $P(1)$. We have two cases: 1) $P(1) = +\Delta p$ or 2) $P(1) = -\Delta p$.

*Case $P(1) = +\Delta p$.* According to the definition of $+\Delta$ either i) $p \in F$ or ii) $\exists r \in R_s[p]$ such that $A(r) = \emptyset$.

i) If $p \in F$, then in $normal(D)$ we have the rules $\to p'$ and $p' \to p$; according to clause 2 of $+\Delta$, $P'(1) = +\Delta p'$ is a proof of $p'$ in $normal(D)$, then we can apply the same clause with respect to the rule $p' \to p$ to derive $P'(2) = +\Delta p$ in $D'$.

ii) The rule $r$ used to derive $p$ has the form $\to p$, and in $R'$ we have we have the rules $\to p'$ and $p' \to p$, so we can repeat the argument for the previous case.

*Case $P(1) = -\Delta p$.* This implies i) $p \notin F$ and ii) $R_s[p] = \emptyset$. In $R'_s$ we have the rule $p' \to p$, but from i) and ii) we know that there is no strict rule for $p'$, and $p'$ cannot be in $F'$, since $F' = \emptyset$ and $p'$ is not in $\Sigma$. Therefore according to $-\Delta$, $P'(1) = -\Delta p'$ is a proof in $normal(D)$. We can now apply the definition of $-\Delta$ with respect to $p$. If $R_{sd}[p] = \emptyset$ then $R'_s[p] = \emptyset$, and trivially $-\Delta p$. Otherwise the only rule for $p$ in $R'$ is $p' \to p$, the only rule for $p$ is $p' \to p$, and all the literals occurring in the antecedent of such a rule have tag $-\Delta$, therefore we can append $P'(2) = -\Delta p$ to $P'(1)$ to obtain a proof of $-\Delta p$ in $D'$.

We have thus proved the inductive base.

*Inductive step $n > 1$.* Let us assume that the property holds up to $n$. We have to consider four cases: 1) $P(n+1) = +\Delta p$, 2) $P(n+1) = -\Delta p$, 3) $P(n+1) = +\partial p$, and $P(n+1) = -\partial p$.

*Case $P(n+1) = +\Delta p$.* We consider only the case different from the analogous case of the inductive base. Here we have to consider a rule $r : A(r) \to p$, where $\forall a_r \in A(r), +\Delta a_r \in P(1..n)$. By inductive hypothesis, each $a_r \in A(r)$ is provable in $normal(D)$. However, by construction, in $D'$ there is only one strict rule for each $a_r$, and such a rule has form $a'_r \to a_r$. Let $P'_r$ be the proof of $a_r$ in $D'$, and $+\Delta a'_r \in P'_r(1..n_r)$. We concatenate the proofs of the $a_r$'s, and we append $+\Delta p'$ and $+\Delta p$. It is immediate to verify that the result is a proof of $p$ in $normal(D)$.

*Case $P(n+1) = -\Delta p$.* Let us assume that the property holds up to $n$ and $-\Delta p \in P(n+1)$. This means that $\forall r \in R_s[p] \exists a \in A(r)$ such that $-\Delta a \in P(1..n)$.

We have two cases:

- If $r \in R_s[p]$, then, in $normal(D)$ we have $r' : A(r)' \to p'$ and $p' \to p$; thus we have to show that $normal(D) \vdash -\Delta p'$. The strict rules for $p$ in $D$ correspond to the strict rules for $p'$ in $normal(D)$. From the hypothesis we know that $p \notin F$ and all strict rules $r$ for $p$ are discarded, then, by inductive hypothesis, the corresponding rules $r'$ are discarded too.

- If $R_{sd}[p] = \emptyset$ then there are no rule for $p$ in $R'$, and trivially $-\Delta p$. If $R_s[p] = \emptyset$ but $R_d[p] \neq \emptyset$, then $p' \to p \in R'$. Since there are no strict rules for $p$ in $D$, there are no strict rules for $p'$ in $D'$, therefore $normal(D) \vdash -\Delta p'$, and so the only rule for $p$ in $normal(D)$ ($p' \to p$) is discarded.

We have thus proved that in each case the strict rules for $p$ are discarded in $normal(D)$, therefore $D' \vdash -\Delta p$.

*Cases* $P(n+1) = +\partial p$ *and* $P(n+1) = -\partial p$. It is enough to notice that the structure (including the superiority relation) of defeasible rules and defeaters in $R'$ is identical with the structure of all rules in $R$. Thus, when deriving a defeasible conclusion concerning a literal from $D$ (i.e., not involving a new proposition), the only difference between $D$ and $normal(D)$ is the presence of rules $p' \to p$, but such rules become relevant only when $+\Delta p'$ is provable, that means that also $+\Delta p$ is provable.

*Case* $\Leftarrow$. or each literal $p \in \Sigma$ the only strict rule for it (if any) in $normal(D)$ is $p' \to p$. Then if $P'(n+1) = +\Delta p$, then $+\Delta p' \in P(1..n)$; and if $P'(n+1) = -\Delta p$, then $-\Delta p' \in P(1..n)$. If we replace each $p'$ with $p$ and each rule $r'$ with $r$ in $P'$, we obtain a proof in $D$. For $+\partial p$ and $-\partial p$ we can repeat the same considerations of the previous case. $\square$

**Proposition 8** *The transformation normal is incremental, but not modular.*

**Proof.** It is immediate to see that for every pair of defeasible theories $D_1, D_2$, $normal(D_1) \cup normal(D_2) = normal(D_1 \cup D_2)$. Consequently $normal$ is incremental.

To see that $normal$ is not modular, consider $D_1 = \{a \to b\}$ and $D_2 = \{\to a\}$. Then $normal(D_1) = \{a \Rightarrow b, a' \Rightarrow b', b' \to b\}$. Clearly $D_1 \cup D_2 \vdash +\Delta b$. However $normal(D_1) \cup D2 \vdash -\Delta b$ since there is no fact $a'$. $\square$

This demonstrates, as promised, that the inverse of Proposition 4 does not hold.

## 5.2 Simulating the Superiority Relation

In this section we show that the superiority relation does not contribute anything to the expressive power of Defeasible Logic. Of course it does allow one to represent information in a more natural way.

We define below a transformation *elim_sup* that eliminates all uses of the superiority relation. For every rule $r$, it introduces two new previously unused positive literals denoted by $inf^+(r)$ and $inf^-(r)$. Intuitively, $inf^+(r)$ and $inf^-(r)$ express that $r$ is overruled by a superior rule.

**Definition 9** *Let* $D = (\emptyset, R, >)$ *be a normal defeasible theory. Let* $\Sigma$ *be the language of* $D$. *Define* $elim\_sup(D) = (\emptyset, R', \emptyset)$, *where*

$$R' = R_s \cup \{\neg inf^+(r_1) \Rightarrow inf^+(r_2),$$
$$\neg inf^-(r_1) \Rightarrow inf^-(r_2) \mid r_1 > r_2\} \cup$$
$$\{A(r) \Rightarrow \neg inf^+(r),$$
$$\neg inf^+(r) \Rightarrow p,$$
$$A(r) \Rightarrow \neg inf^-(r),$$
$$\neg inf^-(r) \Rightarrow p \mid r \in R_d[p]\} \cup$$
$$\{A(r) \Rightarrow \neg inf^-(r),$$
$$\neg inf^-(r) \leadsto p \mid r \in R_{dft}[p]\}$$

*For each* $r$, $inf^+(r)$ *and* $inf^-(r)$ *are new atoms not in* $\Sigma$. *Furthermore all new atoms are distinct.*

A defeasible proof of a literal $p$ consists of three phases. In the first phase either a strict or defeasible rule is put forth in order to support a conclusion $p$; then we consider an attack on this conclusion using the rules for its negation $\sim p$. The attack fails if each rule for $\sim p$ is either discarded (it is possible to prove that part of the antecedent is not defeasibly provable) or if we can provide a stronger counterattack, that is, if there is an applicable strict or defeasible rule stronger than the rule attacking $p$. It is worth noting that defeaters cannot be used in the last phase. For this reason we have introduced two predicate $inf^+(r)$ and $inf^-(r)$ for each rule $r$. Intuitively $\neg inf^+(r)$ means that $r$ is not inferior to any applicable strict or defeasible rule, while $\neg inf^-(r)$ states that $r$ is not inferior to an applicable rule.

Independently, a somewhat similar construction is given in [12] for eliminating priorities among defeasible rules in a credulous abstract argumentation framework. However that transformation does not work properly in Defeasible Logic because of the presence of defeaters. And if defeaters are incorporated in that model, it would still not work properly because defeaters could be used to support positive conclusions trough counterattacks, something prohibited in our logic (and something we consider counterintuitive). In this context it is worth noting that an earlier version of our transformation [2] worked for a different variation of defeasible logic where defeaters could be used for counterattacks.

Before we study the properties of *elim_sup* we provide an example that illustrates how it works.

**Example 3** Let us consider the defeasible theory:

$$
\begin{array}{ll}
r1: \rightarrow gap & \text{Tweety is a genetically altered penguin} \\
r2: gap \rightarrow p & \text{Genetically altered penguins are penguins} \\
r3: p \rightarrow b & \text{Penguins are birds} \\
r4: b \Rightarrow f & \text{Birds usually fly} \\
r5: p \Rightarrow \neg f & \text{Penguins don't fly} \\
r6: gap \leadsto f & \text{Genetically altered penguins might fly} \\
r5 > r4,\ r6 > r5 &
\end{array}
$$

The transformation of the above theory is

$$
\begin{array}{lll}
r4a^+: b \Rightarrow \neg inf^+(r4) & r5a^+: p \Rightarrow \neg inf^+(r5) & r6a: gap \Rightarrow \neg inf^-(r6) \\
r4a^-: b \Rightarrow \neg inf^-(r4) & r5a^-: p \Rightarrow \neg inf^-(r5) & r6b: \neg inf^-(r6) \leadsto f \\
r4b^+: \neg inf^+(r4) \Rightarrow f & r5b^+: \neg inf^+(r5) \Rightarrow \neg f & \\
r4b^+: \neg inf^-(r4) \Rightarrow f & r5b^+: \neg inf^-(r5) \Rightarrow \neg f &
\end{array}
$$

and

$$
\begin{array}{ll}
s1^+: & \neg inf^+(r6) \Rightarrow inf^+(r5) \\
s1^-: & \neg inf^-(r6) \Rightarrow inf^-(r5) \\
s2^+: & \neg inf^+(r5) \Rightarrow inf^+(r4) \\
s2^-: & \neg inf^-(r5) \Rightarrow inf^-(r4)
\end{array}
$$

and

$$
\begin{array}{ll}
r1: & \rightarrow gap \\
r2: & gap \rightarrow p \\
r3: & p \rightarrow b
\end{array}
$$

It is immediate to see that $r_4$ is defeated by $r_5$, and, at the same time, $r_5$ is defeated by $r_6$. However, $r_6$ is a defeater, thus it does not support a conclusion, and, according to clause (2.3.2) of the definition of $-\partial$, it cannot be used to reinstate the conclusion of $r_4$. To represent this fact we have to use two new literals for each rule $r$, i.e., $inf^+(r)$ and $inf^-(r)$. In the presence of defeaters $inf^+$ can be used both to defeat a competing rule or to reinstate the conclusion of a rule with the same conclusion. On the other hand $inf^-$ can be used only to defeat competing rules.

Let us examine the transformed theory. There are no rules against $\neg inf^-(r6)$, so we can derive $+\partial\neg inf^-(r6)$, thus the rule $s1^-$ is applicable; this implies $-\partial\neg inf^-(r5)$, and therefore $r5^-$ is discarded: it cannot be used to support the derivation of $\neg f$, nor prevent the derivation of $f$. Moreover, also $s2^-$ is discarded, then $+\partial inf^-(r4)$ is derivable, thus $r4b^-$ is applicable. If we defined only $inf^-$ literals then $+\partial f$ would be provable. However, we have to consider $inf^+$ literals. There is no rule for $\neg inf^+(r6)$, so $-\partial\neg inf^-(r6)$, $s1^+$ becomes discarded, and we can prove $+\partial\neg inf^+(r5)$. At this point $r5b^+$ is applicable; consequently we have two applicable competing rules, from which we conclude both $-\partial f$, and $-\partial\neg f$.

To formulate the following theorem we need a condition of *distinctness*: Two theories $D_1 = (F_1, R_1, >_1)$ and $D' = (F_2, R_2, >_2)$ are distinct if the rules of one theory do not appear in the superiority relation of the other. Formally: For $i = 1, 2$, if the pair $(r, r')$ is in the relation $>_i$ then neither $r$ nor $r'$ are rules in $R_{3-i}$.

**Theorem 10** *The transformation elim_sup is modular for distinct well-formed normalized defeasible theories. That is, for such theories $D_1$ and $D_2$, $D_1 \cup D_2 \equiv_\Sigma T(D_1) \cup D_2$, where $\Sigma$ is the union of the languages of $D_1$ and $D_2$. Thus elim_sup is also incremental.*

**Proof.** To prove that the transformation *elim_sup* is modular we have to show that $D_1 \cup D_2 \equiv_\Sigma D_1 \cup elim\_sup(D_2)$ and $D_1 \cup D_2 \equiv_\Sigma elim\_sup(D_1) \cup D_2$. We prove only the the first case since the second is symmetrical.

Since $D_1$ and $D_2$ are well-formed theories their superiority relations are acyclic and since $D_1$ and $D_2$ are distinct, so are $>_1$ and $>_2$, moreover they are defined over the rule of $D_1$ and $D_2$. Therefore the superiority relation of the $D_1 \cup D_2$ is acyclic; moreover the superiority relation in $elim\_sup(D_2)$ is empty, therefore in $D_1 \cup elim\_sup(D_2)$ the superiority relation is the superiority relation of $D_1$, which is assumed to be acyclic.

From now on we use $D$ for $D_1 \cup D_2$ and $D'$ for $D_1 \cup elim\_sup(D_2)$.

We prove the theorem by induction on the length of proofs.

*Inductive base $n = 1$.* Suppose the length of a proof $P$ is 1. The only line in $P$, $P(1)$ is either $+\Delta p$ or $-\Delta p$. It is immediate to see that $D$ and $D'$ have the same definite conclusions in the language $\Sigma$ since $R_s = R_{s1} \cup R_{s2}$ $R'_s = R_{s1} \cup elim\_sup(R_{s2})$, $R_{s2} = elim\_sup(R_{s2})$, and the superiority relation does not affect the proof of definite conclusions.

*Inductive step $n > 1$.* We consider only the cases of defeasible conclusions, since, as we have seen in the inductive base, $D$ and $D'$ have the same definite conclusions.

*Case* $D \vdash +\partial p \Rightarrow elim\_sup(D) \vdash +\partial p$. If $+\Delta p \in P(1..n)$ then $D \vdash +\Delta p$; since $D \vdash +\Delta p$ iff $D' \vdash +\Delta p$, and the latter implies $D' \vdash +\partial p$.

Similarly if $+\Delta p \in P'(1..n)$, then $D' \vdash +\Delta p$; since $D \vdash +\Delta p$ iff $D' \vdash +\Delta p$, and the former implies $D \vdash +\partial p$.

We consider the rule $r$ from which $p$ has been derived in $D$, and we consider the rules corresponding to $r$ in $elim\_sup(D)$. We have two cases: 1) $r \in R_1$ and 2) $r \in R_2$.

1) If $r \in R_1$, then $r \in R'$. In this case $\forall a_r \in A(r)$, $+\partial a_r \in P(1..n-1)$, therefore, by inductive hypothesis $D' \vdash +\partial a_r$, thus $r$ is applicable in $D'$.

2) If $r \in R_2$, then we consider the rules corresponding to it in $D'$, namely:

$$r_1 : A(r) \Rightarrow \neg inf^+(r) \quad r_2 : A(r) \Rightarrow \neg inf^-(r) \quad r_3 : \neg inf^+(r) \Rightarrow p \quad r_4 : \neg inf^-(r) \Rightarrow p$$

By inductive hypothesis, both $r_1$ and $r_2$ are applicable, that is, $\forall a_r \in A(r)$, $elim\_sup(D) \vdash +\partial a_r$.

Let us concentrate to what happens when $r \in R_2$. We have two cases a) $r$ is superior, b) $r$ is not superior.

For a) if $r$ is superior then $\neg \exists s : s > r$, then, in $elim\_sup(D)$, $R[inf^+(r)] = \emptyset$, and so is $R[inf^-(r)]$; hence $elim\_sup(D) \vdash +\partial \neg inf^{\pm}(r)$.

For b) if $r$ is not superior we consider the set of rules $S_0$ such that $s > r$ in $D$. Since $D$ is a well-formed theory, $>$ is defined over competing rules; therefore $S_0 \subseteq R[\sim p]$. We consider the transformations of such rules: $s_1 : A(s) \Rightarrow \neg inf^-(s)$, and eventually $s_2 : A(s) \Rightarrow \neg inf^+(s)$ if $s$ is not a defeater. The translation of the instances $s > r$ consists of the rules

$$\neg inf^+(s) \Rightarrow inf^+(s) \qquad \neg inf^-(s) \Rightarrow inf^-(s)$$

If $s \in S_0$ is discarded in $D$, i.e., $\exists a_s \in A(s) : D \vdash -\partial a_s$, by inductive hypothesis so are $s_1$ and $s_2$ in $elim\_sup(D)$, thus they cannot be used to block the derivation of $\neg inf^{\pm}(r)$ in $elim\_sup(D)$.

If $s \in S_0$ is applicable, then according to clause (2.3.2) of the definition of $+\partial$, $\exists t \in R_{sd}[p]$ such that $\forall a_t \in A(t)$, $D \vdash +\partial a_t$ and $t > s$. Again, by inductive hypothesis $elim\_sup(D) \vdash +\partial a_t$. We have to examine two cases: i) $t$ is superior ii) $t$ is not superior. For i) we can repeat the reasoning we have done in a).

For ii) if $t$ is not superior, then we consider the set $S_1 = \{s' : s' > t\}$. Since $D$ is a well-formed theory, $>$ is defined over competing rules; therefore $S_1 \subseteq R[\sim p]$, moreover $S_1 \subset S_0$, since $>$ is acyclic. Therefore we can repeat $n$ times the same reasoning until we arrive at a rule $t' \in R_{sd}[p]$, which is superior and applicable. Hence $\forall a_{t'} \in A(t'), D \vdash +\partial a_{t'}$; by inductive hypothesis, $elim\_sup(D) \vdash +\partial a_{t'}$. The rules corresponding to $t'$ in $elim\_sup$ are

$$t'_1 : A(t') \Rightarrow \neg inf^+(t') \qquad t'_2 : A(t') \Rightarrow \neg inf^-(t')$$

Therefore $elim\_sup(D) \vdash +\partial \neg inf^{\pm}(t')$. According to clause (2.3.2) of the definition of $+\partial$. Moreover from the superiority relation $t' > s'$ for some rule $s' \in S_n$, we have the rules

$$\neg inf^+(t') \Rightarrow inf^+(s') \qquad \neg inf^-(t') \Rightarrow inf^-(s');$$

hence $elim\_sup(D) \vdash -\partial inf^{\pm}(s')$. We can repeat backward the steps leading to $S_n$, and we can conclude $elim\_sup(D) \vdash -\partial inf^{\pm}(s)$, and then $elim\_sup(D) \vdash +\partial \neg inf^{\pm}(r)$. We have thus proved that both $r_3$ and $r_4$ are applicable in $D'$.

At this point to prove $+\partial p$, we have to show that clause (2.3) of the definition of $+\partial$ is satisfied. To this end, we analyse two cases: we consider a rule $s \in R[\ p]$: i) $s \in R_1$ or ii) $s \in R_2$.

i) If $s \in R_1$, then $s \in R'$. If it is discarded in $D$, then, by inductive hypothesis, it is discarded in $D'$ too. Otherwise if $s$ satisfies clause (2.3.2), we consider a rule $t$ that defeats $s$. The superiority relations of $D_1$ and $D_2$ are disjoint, thus no rule in $D_2$ is superior to a rule in $D_1$. Moreover the superiority relation of $D'$ is that of $D_1$, thus $t \in R_1$, and therefore $t \in R'$. By inductive hypothesis $t$ is applicable in $D'$, hence, in this case $+\partial p$ is provable in $elim\_sup(D)$

ii) If $s$ is discarded, then, by inductive hypothesis, the rules corresponding to $s$ in $elim\_sup(D)$, $s_1 : A(s) \Rightarrow \neg inf^-(s)$ and, eventually, $s_2 : A(s) \Rightarrow \neg inf^+(s)$ are discarded too; therefore $elim\_sup(D) \vdash -\partial \neg inf^{\pm}(s)$, from which we infer that the rule $s_3 : \neg inf^-(s) \rightsquigarrow \sim p$, if

$s \in R_{dft}[\sim p]$, or the rules

$$s_3 : \neg inf^-(s) \Rightarrow \sim p \qquad\qquad s_4 : \neg inf^+(s) \Rightarrow \sim p,$$

if $s \in R_{sd}[p]$, are discarded. Otherwise, if $s$ satisfies clause (2.3.2), we can repeat the same argument of b) to prove that $elim\_sup(D) \vdash -\partial \neg inf^\pm(s)$.

In both cases we have proved that the rules for $\sim p$ are discarded, and therefore, since there is an applicable rule for $p$, $elim\_sup(D) \vdash +\partial p$.

*Case $elim\_sup(D) \vdash +\partial p \Rightarrow D \vdash +\partial p$.* If $D' \vdash +\partial p$ because of $D' \vdash +\Delta p$, then we have already proved that $D \vdash +\Delta p$, hence $D \vdash +\partial p$. Otherwise we have to consider the form of the applicable rule used to justify the derivation of $+\partial p$ in $D'$. We have two cases 1) $r : A(r) \Rightarrow p$ 2) $r : \neg inf^\pm(r) \Rightarrow p$. In the first case $r$ corresponds to itself in $D'$; by inductive hypothesis $r$ is applicable in $D$ too. In the second case, according to clause (2.1) of $+\partial p$, it is required that either

$$elim\_sup(D) \vdash +\partial \neg inf^+(r) \qquad \text{or} \qquad elim\_sup(D) \vdash +\partial \neg inf^-(r)$$

The rules for $\neg inf^\pm$ in $elim\_sup(D)$ have form

$$r_3 : A(r) \Rightarrow \neg inf^+(r) \qquad\qquad r_4 : A(r) \Rightarrow \neg inf^-(r)$$

Again, by clause (2.3) of $+\partial$, $\forall a_r \in A(r)$, $elim\_sup(D) \vdash +\partial a_r$. By inductive hypothesis $D \vdash +\partial a_r$. The rules $r_1$, $r_2$, $r_3$, and $r_4$ correspond to rule $r$ in $D$, hence $r$ is applicable in $D$.

We have to consider now the rule for $\sim p$ Similarly, we have two cases: 1) The rule for $\sim p$ corresponding to rule in $R_1$, which are the same in both $D$ and $D'$ and 2) the rules for $\sim p$ with form

$$s_1 : \neg inf^-(s) \leadsto \sim p$$

if $s \in R_{dft}[\sim p]$, otherwise they have form

$$s_1 : \neg inf^-(s) \Rightarrow \sim p \qquad\qquad s_2 : \neg inf^+(s) \Rightarrow \sim p$$

corresponding to rules in $R_2$.

If a rule $s$ for $\sim p$ corresponding to a rule in $R_1$ is discarded in $D'$, so is in $D'$; if it is applicable in $D'$ so is in $D$, and there must be an applicable rule $t$ for $p$ such that $t > s$ Since $>_1$ and $>_2$ are disjoint and $>'=>_1$; therefore $t$ also is in $R_1$.

For the same reason as above no rule of the form $\neg inf^\pm(s) \Rightarrow \sim p$ is inferior to any other rule for $p$, therefore all such rules must be discarded, that is, $elim\_sup(D) \vdash -\partial \neg inf^\pm(s)$. The rules for $\neg inf^\pm(s)$ are

$$s_3 : A(s) \Rightarrow \neg inf^-(s) \qquad\qquad s_4 : A(s) \Rightarrow \neg inf^+(s) \text{ if } s \text{ is not a defeater.}$$

Now $elim\_sup(D) \vdash -\partial \neg inf^\pm(s)$ iff 1) $R[\neg inf^\pm] = \emptyset$, this is the case for $\neg inf^+(s)$, if $s$ is a defeater; or 2) the rules for $\neg inf^\pm(s)$ are discarded; or 3) the rules for $inf^\pm(s)$ are supported. Case 1) is immediate, so we consider 2) and 3).

For 2) The rules for $\neg inf^\pm(s)$ iff $\exists a_s \in A(s)$ such that $elim\_sup(D) \vdash -\partial a_s$. By construction the rules for $\neg inf^\pm(s)$ correspond to $s$, and they have the same body; thus, by inductive hypothesis, $D \vdash -\partial a_s$, and therefore $s$ is discarded.

For 3) The rules for $inf^\pm(s)$ have form $\neg inf^\pm(t) \Rightarrow inf^\pm(s)$, and they correspond to instances of $t > s$ in $D$. Since they are applicable, we have $elim\_sup(D) \vdash +\partial \neg inf^\pm(t)$. The superiority relation in $D$ is defined over competing rules, and since $+\partial \neg inf^+(t)$ can be proved, $t \in R_{sd}[p]$ in $D$. We know that if $elim\_sup(D) \vdash +\partial \neg inf^\pm(t)$ then the rules for $\neg inf^\pm(t)$ must be applicable. Such rules have form

$$t_1 : A(t) \Rightarrow \neg inf^+(t) \qquad t_2 : A(t) \Rightarrow \neg inf^-(t)$$

Thus, $\forall a_t \in A(t)$, $elim\_sup(D) \vdash +\partial a_t$. By inductive hypothesis $D \vdash +\partial a_t$ for all $a_t$ in $A(t)$.

From 2) and 3) we can conclude that in $D$ there is an applicable rule $r$ for $p$, such that for every rule $s$ for $\sim p$, either $s$ is discarded or there exists an applicable rule $t$ for $p$ such that $t$ is stronger than $s$, therefore $D \vdash +\partial p$.

*Case $D \vdash -\partial p \Rightarrow elim\_sup(D) \vdash -\partial p$.* If $P(n) = -\partial p$ because $+\Delta \sim p \in P(1..n-1)$, then, by inductive hypothesis $D' \vdash +\Delta \sim p$, and therefore $D' \vdash -\partial p$.

Let us consider now the remaining cases. Let $r$ be a rule in $R_{sd}[p]$. If $r$ is discarded, then, if $r \in R_1$, then, by construction, $r$ is also in $R'$, and by inductive hypothesis is discarded in $D'$. If $r \in R_2$, then in $R'$ we have the rules

$$r_1^+ : A(r) \Rightarrow \neg inf^+(r) \qquad r_1^- : A(r) \Rightarrow \neg inf^-(r)$$
$$r_2^+ : \neg inf^+(r) \Rightarrow p \qquad r_2^- : \neg inf^-(r) \Rightarrow p$$

We know that $r$ is discarded in $D$, thus $\exists a_r \in A(r)$ such that $D \vdash -\partial a_r$, then, by inductive hypothesis, $elim\_sup(D) \vdash -\partial a_r$. The only rules for $\neg inf^\pm(r)$ are $r_1^\pm$; this implies that $elim\_sup(D) \vdash -\partial \neg inf^\pm(r)$, and hence the rules $r_2^\pm$ are discarded.

If $-\partial p$ can be proved because its proof satisfied clause 2.3, then there exists a rule $s$ such that 1) $s$ is applicable and $s$ is not inferior to any applicable rule for $p$.

Again if $s \in R_1$, then $s \in R'$. Since $R_1$ and $R_2$ are disjoint no rule in $R_1$ is inferior to a rule in $R_2$. Moreover the superiority relation of $D'$ is that of $R_1$, therefore by construction and inductive hypothesis $s$ is applicable in $D'$ and it is not inferior to any applicable rule in $D'$.

If $s \in R_2$, we consider the rule corresponding to it in $R'$:

$$s_1^\pm : \neg inf^\pm(s) \Rightarrow \sim p \qquad s_2^\pm : A(s) \Rightarrow \neg inf^\pm(s)$$

if $s \in R_d$ and

$$s_1 : \neg inf^-(s) \rightsquigarrow \sim p \qquad s_2 : A(s) \Rightarrow \neg inf^-(s)$$

if $s$ is a defeater. In both cases, by inductive hypothesis the rules for $\neg inf^\pm(s)$ are applicable. Let us consider the rules for $inf^\pm(s)$; they correspond to instances of the superiority relation of $D_2$, $t > s$, where $t \in R[p]$ since $D_2$ is a well-formed theory. Such rules have form

$$\neg inf^\pm \Rightarrow inf^\pm(s)$$

However, according to clause $(-\partial 2.3.1)$ $\forall t \in R_{sd}[p]$ either a) $\exists a_t \in A(t)$ such that $D \vdash -\partial a_t$ or $t \not> s$. From be we obtain that the rules $\neg inf^\pm(t) \Rightarrow inf^\pm(s)$ do not exist, while from a), by inductive hypothesis, we get that the rules $\neg inf^\pm \Rightarrow inf^\pm(s)$ are discarded: the only rules for $\neg inf^\pm(t)$ are $A(r) \Rightarrow \neg inf^\pm(t)$, but, by inductive hypothesis, they are discarded. Therefore $D' \vdash +\partial \neg inf^\pm(s)$.

We have to consider that a defeater $t$ is superior to $s$, but from the defeater we have no rules for $\neg inf^+(t)$, and therefore $D' \vdash -\partial \neg inf^+(t)$, thus the rule $\neg inf^+ \Rightarrow inf^+(s)$ is discarded. Hence in all cases we are able to prove $D' \vdash +\partial \neg inf^+(s)$. That means that the rules $\neg inf^+(s) \Rightarrow \sim p$ or $\neg inf^+ \rightsquigarrow \sim p$ are applicable in $D'$, and therefore also in this case $D' \vdash -\partial p$.

*Case elim_sup(D) $\vdash -\partial p \Rightarrow D \vdash -\partial p$*. We consider the form of the rules for $p$ in $R'_{sd}[p]$. We have two possibilities:

1) $A(r) \Rightarrow p$ 　　　　　　　　　 2) $\neg inf^{\pm}(r) \Rightarrow p$

In the first case the rule belongs to $R_1$, and in $D'$ we have the same rule; in the other case the rules correspond to $r : A(r) \Rightarrow p$ in $D$, and in $D'$ we have also $A(r) \Rightarrow \neg inf^{\pm}(r)$.

By hypothesis we have a proof $P$ in $D'$ where the last line in $-\partial p$. If a rule $r \in R'_s[p]$ is discarded, then if it has form 1) then, by inductive hypothesis, the same rule is discarded in $D$. Let us see what happens in the other case. Here we have that $D' \vdash -\partial \neg inf^{\pm}(r)$. The rules for $\neg inf^{\pm}(r)$ are

$$A(r) \Rightarrow \neg inf^{\pm}(r)$$

and those for $inf^{\pm}(r)$ have form

$$\neg inf^{\pm}(s) \Rightarrow inf^{\pm}(r)$$

Such rules, if any, correspond to instances of the superiority relation $s > r$ in $D$. Thus to prove $-\partial \neg inf^{\pm}(r)$ we have to prove either that the rule for it is discarded, or that a rule for $inf^{\pm}(r)$ is applicable. If it is discarded then, by inductive hypothesis the corresponding rule in $D$ is discarded too. If the rule for $inf^{\pm}$ is applicable the $D' \vdash +\partial \neg inf^{\pm}(s)$.[2]

The rules for $\neg inf^{\pm}(s)$ have form $A(s) \Rightarrow \neg inf^{\pm}(s)$. Thus, since $+\partial \neg inf^{\pm}(s)$ is provable in $D'$ we have that $\forall a_s \in A(s), D' \vdash +\partial a_s$, then by inductive hypothesis the corresponding rule $s$ is applicable in $D$. Moreover since $D_2$ is a well-formed theory $s \in R[\sim p]$.

We have now to prove that $s$ is not inferior to any applicable rule for $p$ in $D$. From the hypothesis we know that $D' \vdash +\neg inf^{\pm}(s)$; this implies that the rules for $inf^{\pm}(s)$, if any, are discarded. Since the superiority relation of $D'$ is the superiority relation of $D_1$, there is no superiority relation on rules for $inf$ literals.

If there are no rules for $inf^{\pm}(s)$, then there is no rule $t$ in $D$ such that $t > s$. Otherwise the rules for $inf^{\pm}(s)$ have form

$$\neg inf^{\pm}(t) \Rightarrow inf^{\pm}(s)$$

Thus, if they are discarded, $D' \vdash -\partial \neg inf^{\pm}(t)$. The rules for $\neg inf^{\pm}(t)$ are $A(t) \Rightarrow \neg inf^{\pm}(t)$, while those for $inf^{\pm}(t)$ have form $\neg inf^{\pm}(s') \Rightarrow inf^{\pm}(t)$.

Since we have a proof of $+\partial \neg inf^{\pm}(s)$, and proofs are finite sequences of nodes, we can repeat the cycle above for a finite number of times until we arrive at a point where we can prove $+\partial \neg inf^{\pm}(s^*)$, and either we have no rules $inf^{\pm}(s^*)$, or all rules for it (i.e., $\neg inf^{\pm}(t^*) \Rightarrow inf^{\pm}(s^*)$) are discarded or there are no rules for $inf^{pm}(t^*)$. This implies that the corresponding rule $s^*$ in $D$ is in $R[\sim p]$, and by inductive hypothesis it is applicable, and $s^*$ is not inferior to an applicable rule for $p$.

If a rule for $p$ is applicable in $D'$, so is the corresponding rule in $D$, and we have to show that such a rule is defeated. To this end we consider the rules for $\sim p$. As usual we have two cases[3]:

$s : A(s) \hookrightarrow \sim p$ 　　　　　　　　　 $s' : \neg inf^{\pm}(s) \hookrightarrow \sim p$

---

[2]Notice that the superiority relation is empty for the *inf* symbols, therefore clause 2.3.2 is vacuously satisfied.
[3]Here $\hookrightarrow$ stands for either $\Rightarrow$, or $\leadsto$.

In the first case $s \in R_1$, and by inductive hypothesis and the considerations about the superiority relation we have made above, if $s$ is applicable and it is not inferior to any rule for $p$ in $D$, so is in $D'$.

In the second case we have already shown that there exists a rule $s^*$ such that $s^* \in R[\sim p]$, and it is not inferior to any applicable rule for $p$ in $D$. Therefore in both cases $D \vdash -\partial p$.

Incrementality follows immediately from Proposition 4. □

The following result follows immediately from Proposition 4.

**Corollary 11** *The transformation elim_sup is correct for well-formed normalized theories.*

Theorem 10 imposes two conditions under which *elim_sup* is modular: acyclicity of the superiority relation, and distinctness (of $D_1$ and $D_2$). In the following we show that these conditions are not needed for the modularity of *elim_sup* only, but they are necessary conditions for the existence of any modular transformation that empties the superiority relation. The first result shows that acyclicity is a necessary condition for the incrementality (and therefore modularity) of such a transformation.

**Theorem 12** *Let $D = (F, R, >)$ be a (possibly cyclic) defeasible theory. Then in general there is no correct and incremental transformation $T$ such that $T(D) = (F', R', \emptyset)$. It follows that there is no modular such transformation either.*

**Proof.** Let $D_1$ and $D_2$ be a partition of $D$, where

$$D_1 = (\emptyset, r_1 : \Rightarrow p, r_1 > r_2) \qquad D_2 = (\emptyset, r_2 : \Rightarrow \neg p, r_2 > r_1)$$

Then, the theory $D$ is cyclic and both $+\partial p$ and $+\partial \neg p$ are provable. On the other hand $T(D_1) \cup T(D_2)$ is acyclic (the superiority relation is empty); by consistency $+\partial p$ and $+\partial \neg p$ are both provable only if (1) the theory is cyclic or (2) $+\Delta p$ and $+\Delta \neg p$ are both provable. However $-\Delta p$ and $-\Delta \neg p$ are provable in $D$ so if the transformation $T$ is correct $-\Delta p$ and $-\Delta \neg p$ should be provable in $T(D_1) \cup T(D_2)$, and therefore $+\partial p$ and $+\partial \neg p$ are not provable simultaneously. Thus $T(D_1) \cup T(D_1)$ is not equivalent with respect to $\Sigma$ to $D = D_1 \cup D_2$, thus contradicting the correctness of $T$.

Proposition 4 tells us that there can be no such modular transformation either. □

Our next result states that even for acyclic defeasible theories there is no modular transformation for simulating the superiority relation.

**Theorem 13** *Let $D = (F, R, >)$ be a defeasible theory. Then in general there is no modular transformation $T$ such that $T(D) = (F', R', \emptyset)$.*

**Proof.** Let us consider the defeasible theory $D$ consisting of

$$r_1 : \Rightarrow p$$
$$r_2 : \Rightarrow \neg p$$
$$r_1 > r_2$$

We partition $D$ into $D_1 = \{r_1 : \Rightarrow p, r_2 : \Rightarrow \neg p\}$ and $D_2 = \{r_1 > r_2\}$. Let us suppose that a modular transformation $T$ removing the superiority relation exists. According to the definition of modularity we have $D \equiv_\Sigma D_1 \cup T(D_2)$. It is easy to see that $D \vdash +\partial p$. Since $D_1 \cup T(D_2)$ contains an applicable rule for $\neg p$ (i.e. $r_2$) and the superiority relation is empty, $D_1 \cup T(D_2) \vdash -\partial p$. But then $D \not\equiv_\Sigma D_1 \cup T(D_2)$, which contradicts our assumption. □

## 5.3 Simulating the Defeaters

Similarly to what we have done in the previous section we show that the defeaters do not contribute to the expressivity of defeasible logic and that they can be simulated by means of strict and defeasible rules.

In the following we present a transformation *elim_dft* that transforms every defeasible theory into an equivalent defeasible theory without defeaters. To this end for every atom $p$ occurring as the consequent of either a defeasible rule or a defeater we introduce two new atoms $p^+$ and $p^-$.

**Definition 14** Let $D = (F, R, >)$ be a defeasible theory, and let $\Sigma$ be the language of $D$. Define $elim\_dft(D) = (F, R', >')$ where:

$$R' = \bigcup_{r \in R} elim\_dft(r)$$

and

$$elim\_dft(r) = \begin{cases} \{r^+ : A(r) \to p^+, \ r^- : A(r) \to \neg p^-, \ r : p^+ \to p\} & r \in R_s[p] \\ \{r^- : A(r) \to p^-, \ r^+ : A(r) \to \neg p^+, \ r : p^- \to \neg p\} & r \in R_s[\neg p] \\ \{r^+ : A(r) \Rightarrow p^+, \ r^- : A(r) \Rightarrow \neg p^-, \ r : p^+ \Rightarrow p\} & r \in R_d[p] \\ \{r^- : A(r) \Rightarrow p^-, \ r^+ : A(r) \Rightarrow \neg p^+, \ r : p^- \Rightarrow \neg p\} & r \in R_d[\neg p] \\ \{r : A(r) \Rightarrow \neg p^-\} & r \in R_{dft}[p] \\ \{r : A(r) \Rightarrow \neg p^+\} & r \in R_{dft}[\neg p] \end{cases}$$

The superiority relation $>'$ is defined by the following condition

$$\forall r', s' \in R' \ (r' >' s' \Leftrightarrow \exists r, s \in R \ r' \in elim\_dft(r), s' \in elim\_dft(s), r' > s, r', s' \text{ are conflicting})$$

For each atom $p \in \Sigma$, $p^+$ and $p^-$ are new atoms, that is they do not appear in $\Sigma$. Furthermore all new atoms generated are distinct.

As we have seen defeaters neither directly support conclusions nor can they be used in the counterattack phase; thus to simulate them we have to introduce two new atoms $p^+$ and $p^-$ for each atom $p$. Intuitively the first ($p^+$) is used to prove $p$, and the second ($p^-$) to block it, thus they roughly correspond, respectively, to the literals $p$ and $\neg p$. This is way a defeater $A(r) \rightsquigarrow p$ is translated into $A(r) \Rightarrow \neg p^-$. It cannot support $p$, but it can be used to attack $\neg p$. On the other hand defeasible rules do not suffer from this drawback, they both support, attack, and counterattack conclusions, so their translation is twofold, and we replace each defeasible rule $A(r) \Rightarrow p$ with the rules: $A(r) \Rightarrow p^+$, $A(r) \Rightarrow \neg p^-$, and $p^+ \Rightarrow p$. The first and the third rule together support the derivation of $p$, while the second attack $\neg p$.

**Example 4** Let us consider the defeasible theory of example 3. We apply to it the transformation *elim_dft*, obtaining the rules

$$
\begin{array}{lll}
r1^+ : \to gap^+ & r1^- : \to \neg gap^- & r1 : gap^+ \to gap \\
r2^+ : gap \to p^+ & r1^- : gap \to \neg p^- & r1 : p^+ \to p \\
r3^+ : p \to b^+ & r3^- : p \to \neg b^- & r3 : b^+ \to b \\
r4^+ : b \Rightarrow f^+ & r4^- : b \Rightarrow \neg f^- & r4 : f^+ \to f \\
r5^+ : p \Rightarrow \neg f^+ & r5^- : p \to f^- & r5 : f^- \to \neg f \\
r6 \ : gap \Rightarrow \neg f^- & &
\end{array}
$$

and the superiority relation

$$r6 > r5^- \qquad r5 > r4 \qquad r5^+ > r4^- \qquad r5^- > r4^+$$

**Theorem 15** *The transformation elim_dft is correct.*

**Proof.**

We prove the theorem by induction on the length of proofs. In what follows we use $P$ to denote a proof in $D$, and $P'$ for a proof in $elim\_dft(D)$.

*Inductive base* $n = 1$. Suppose the length of a proof $P$ is 1. The only line in $P$, $P(1)$ is either $+\Delta p$ or $-\Delta p$.

*Case If* $P(1) = +\Delta p$, *then* $elim\_dft(D) \vdash +\Delta p$. According to the definition of $+\Delta$ either i) $p \in F$ or ii) $\exists r \in R_s[p]$ such that $A(r) = \emptyset$.

i) If $p \in F$, then it suffices to notice that $D$ and $elim\_dft(D)$ have the same set of facts.

ii) The rule $r$ used to derive $p$ has the form $r :\to p$, and in $R'$ we have the rules $r^+ :\to p^+$ and $r : p^+ \to p$. Therefore it is immediate to see that the sequence of tagged literals $+\Delta p^+$ and $+\Delta p$, is a proof of $+\Delta p$ in $elim\_dft(D)$.

*Case If* $P(1) = -\Delta p$, *then* $elim\_dft(D) \vdash -\Delta p$. According to the definition $-\Delta$ both $p \notin F$ and $R_s[p] = \emptyset$. By construction of $elim\_dft(D)$ $p \notin F'$ and $R'_s[p] = \emptyset$; therefore $elim\_dft(D) \vdash -\Delta p$.

*Case If* $P'(1) = +\Delta p$, *then* $D \vdash +\Delta p$. All strict rules (if any) for $p$ in $elim\_dft(D)$ have form $p^+ \to p$, thus $p$ has to be a fact. The set of facts in $elim\_dft(D)$ and $D$ is the same, thus $D \vdash +\Delta p$.

*Case If* $P'(1) = -\Delta p$, *then* $D \vdash -\Delta p$. $p \notin F'$ and $R'_s[p] = \emptyset$, but by construction of $elim\_dft(D)$ this is possible only if $p \notin F$ and $R_s[p] = \emptyset$. Therefore $D \vdash -\Delta p$.

*Inductive base* $n > 1$. We assume that the theorem holds for proof with less than $n + 1$ lines. We prove only the cases different from the inductive base.

*Case If* $P(n + 1) = +\Delta p$, *then* $elim\_dft(D) \vdash +\Delta p$. In this case we have a rule $r \in R_s[p]$ such that $\forall a \in A(r), +\Delta a \in P(1..n)$. In $elim\_dft(r)$ we have the rules

$$r^+ : A(r) \to p^+ \qquad\qquad r : p^+ \to p$$

By inductive hypothesis $r^+$ is applicable, thus we can derive $+\Delta p^+$, which implies that $r$ is applicable too; therefore $elim\_dft(D) \vdash +\Delta p$.

*Case If* $P(n + 1) = -\Delta p$, *then* $elim\_dft(D) \vdash -\Delta p$. Each rule for $p$ is discarded, that is, there exists a literal $a$ such that $-\Delta a \in p(1..n)$. By inductive hypothesis $elim\_dft(D) \vdash -\Delta a$. By construction in $elim\_dft(D)$ we a strict rule for $p^+$ ($r^+$) for each strict rule for $p$ ($r$) in $d$; moreover $r^+$ and $r$ have the same antecedent, therefore $elim\_dft(D) \vdash -\Delta p^+$. The rules for $p$ in $elim\_dft(D)$ have all form $p^+ \to p$, therefore they are all discarded, thus $elim\_dft(D) \vdash -\Delta p$.

*Case If* $P'(n + 1) = +\Delta p$, *then* $D \vdash +\Delta p$. In $elim\_dft(D)$ each rule for $p$ has form $p^+ \to p$; then $+\Delta p^+ \in P(1..n)$. This means there is an applicable strict rule $r^+$ for $p^+$. By construction $r^+$ corresponds to a strict rule $r$ for $p$ in $D$ with the same antecedent; by inductive hypothesis $r$ is applicable too, consequently $D \vdash +\Delta p$.

*Case If* $P'(n+1) = -\Delta p$, *then* $D \vdash -\Delta p$. The rules for $p$ have form $p^+ \to p$, therefore $-\Delta p^+ \in P'(1..n)$. This means that every rule for $p^+$ is discarded. By construction of $elim\_dft(D)$ there is a one to one correspondence between the rule for $p^+$ in $elim\_dft(D)$ and those for $p$ in $D$; moreover corresponding rules have the same antecedent. By inductive hypothesis each strict rule for $p$ in $D$, is discarded, thus $D \vdash -\Delta p$.

Before proving the remaining cases we need some preliminary results. By construction the rules in $D$ and $elim\_dft(D)$ are closely related. Defeasible rules for $p^+$ and $p^-$ are the same as the defeasible rules in $D$ for $p$ and $\neg p$, respectively, apart from the head of the rules. Similarly, defeasible rules for $\neg p^+$ and $\neg p^-$ are the same as the non-strict rules (defeasible rules and defeaters) in $D$ for $\neg p$ and $p$, respectively, apart from head and type of arrow. Because of the direct link between $p^+$ and $p$, and $p^-$ and $\neg p$, we can use induction to establish direct relationships between provability in $D$ of $p$ and $\neg p$, and provability in $elim\_dft(D)$ of $p^+$ and $p^-$:

*Auxiliary Lemma*

1. a proof of $\pm \partial p$ in $D$ can be translated straightforwardly into a proof of $\pm \partial p^+$ in $elim\_dft(D)$;

2. a proof of $\pm \partial \neg p$ in $D$ can be translated straightforwardly into a proof of $\pm \partial p^-$ in $elim\_dft(D)$;

3. a proof of $\pm \partial p^+$ in $elim\_dft(D)$ can be translated straightforwardly into a proof of $\pm \partial p$ in $D$;

4. a proof of $\pm \partial p^-$ in $elim\_dft(D)$ can be translated straightforwardly into a proof of $\pm \partial \neg p$ in $D$.

*Proof of Auxiliary Lemma.* Because the proofs of these results all have the same structure, we will present only one proof (claims 1 and 3 for $+\partial p$). The interested reader should have no problem in adapting it to prove the remaining results.

$$D \vdash +\partial p \text{ iff } elim\_dft(D) \vdash +\partial p^+$$

Suppose this result holds if the proof of $+\partial p$ takes fewer than $n$ lines. We write $D \vdash_n c$ if conclusion $c$ has a proof with fewer than $n$ lines with defeasible tags.

$D \vdash +\partial q$ iff

(1) $D \vdash +\Delta q$ or
(2) (2.1) $\exists r \in R_{sd}[q] \forall a \in A(r) : D \vdash_n +\partial a$ and
    (2.2) $D \vdash -\Delta \sim q$ and
    (2.3) $\forall s \in R[\sim q]$ either
        (2.3.1) $\exists a \in A(s) : D \vdash_n -\partial a$ or
        (2.3.2) $\exists t \in R_{sd}[q]$ such that
            $\forall a \in A(t) : D \vdash_n +\partial a$ and $t > s$

Now, using the induction hypothesis, the fact that $D$ and $elim\_dft(D)$ have the same strict conclusions, and the close relationship between rules from $D$ and $elim\_dft(D)$, the above implies

(1) $elim\_dft(D) \vdash +\Delta q^+$ or
(2) (2.1) $\exists r \in R'_{sd}[q^+] \forall a \in A(r) : elim\_dft(D) \vdash +\partial a$ and
    (2.2) $elim\_dft(D) \vdash -\Delta \sim q^+$ and
    (2.3) $\forall s \in R'[\sim q^+]$ either
        (2.3.1) $\exists a \in A(s) : elim\_dft(D) \vdash -\partial a$ or
        (2.3.2) $\exists t \in R'_{sd}[q^+]$ such that
            $\forall a \in A(t) : elim\_dft(D) \vdash +\partial a$ and $t > s$

and this, of course, is equivalent to $elim\_dft(D) \vdash +\partial q^+$. By induction, this direction of the result holds for proofs of arbitrary length.

In the other direction, we use induction on the defeasible length of proofs in $elim\_dft(D)$. If

(1) $elim\_dft(D) \vdash +\Delta q^+$ or
(2) (2.1) $\exists r \in R'_{sd}[q^+] \forall a \in A(r) : elim\_dft(D) \vdash_n +\partial a$ and
  (2.2) $elim\_dft(D) \vdash -\Delta \sim q^+$ and
  (2.3) $\forall s \in R'[\sim q^+]$ either
    (2.3.1) $\exists a \in A(s) : elim\_dft(D) \vdash_n -\partial a$ or
    (2.3.2) $\exists t \in R'_{sd}[q^+]$ such that
      $\forall a \in A(t) : elim\_dft(D) \vdash_n +\partial a$ and $t > s$

then, in the same way as above,

(1) $D \vdash +\Delta q$ or
(2) (2.1) $\exists r \in R_{sd}[q] \forall a \in A(r) : D \vdash +\partial a$ and
  (2.2) $D \vdash -\Delta \sim q$ and
  (2.3) $\forall s \in R[\sim q]$ either
    (2.3.1) $\exists a \in A(s) : D \vdash -\partial a$ or
    (2.3.2) $\exists t \in R_{sd}[q]$ such that
      $\forall a \in A(t) : D \vdash +\partial a$ and $t > s$

and so $D \vdash +\partial q$.

By induction, this direction holds generally.

*End of Proof of Auxiliary Lemma.*

Now we return to the main proof. It follows immediately from the fact that the only rules for $p$ in $elim\_dft(D)$ are $p^+ \Rightarrow p$ that for every literal $p$:

- If $D \vdash -\partial p$ then $elim\_dft(D) \vdash -\partial p$

- If $elim\_dft(D) \vdash +\partial p$ then $D \vdash +\partial p$

Using the inference rule for $-\partial$ and the specific rules in $elim\_dft(D)$ we have:

If $elim\_dft(D) \vdash -\partial p$ then $elim\_dft(D) \vdash -\Delta p$ and either $elim\_dft(D) \vdash -\partial p^+$ or $[elim\_dft(D) \vdash +\partial p^-$ and either $elim\_dft(D) \vdash -\partial p^+$ or $\exists s' \in R'_{sd}[\sim p] \forall t' \in R'_{sd}[p]\ t' \not> s'$ ].

Using the above properties, and the definition of $>'$ in $elim\_dft$, this is equivalent to

$D \vdash -\Delta p$ and either $D \vdash -\partial p$ or $[D \vdash +\partial \sim p$ and either $D \vdash -\partial p$ or $\exists s \in R_{sd}[\neg p] \forall t \in R_{sd}[p]\ t \not> s$ ]

and this implies $D \vdash -\partial p$. Thus we have established that if $elim\_dft(D) \vdash -\partial p$ then $D \vdash -\partial p$.

If $D \vdash +\partial p$ then $elim\_dft(D) \vdash +\partial p^+$, by the above properties. Hence, there is an applicable rule for $p$ in $elim\_dft(D)$. If $D \vdash +\partial p$ then $D \vdash +\Delta p$ or there is an applicable rule for $p$ in $D$ and either $D \vdash -\partial \sim p$ or for every applicable rule $s$ for $\sim p$ in $D$, there is an applicable defeasible or strict rule $t$ for $p$ in $D$ such that $t > s$. (Here a rule is applicable if its body can be proved defeasibly.)

Thus $elim\_dft(D) \vdash +\Delta p$ or $elim\_dft(D) \vdash -\partial p^-$ or for every applicable rule $s'$ for $\sim p$ in $elim\_dft(D)$, there is an applicable defeasible or strict rule $t'$ for $p$ in $elim\_dft(D)$ such that $t' >' s'$.

In the first case, clearly $elim\_dft(D) \vdash +\partial p$. In the second case, $elim\_dft(D) \vdash -\partial \sim p$. In both this and the third case, it now follows that $elim\_dft(D) \vdash +\partial p$ since, as observed above, there is an applicable rule for $p$ in $elim\_dft(D)$.

Thus we have established that if $D \vdash +\partial p$ then $elim\_dft(D) \vdash +\partial p$. $\square$

**Proposition 16** *The transformation elim_dft is incremental, but not modular.*

**Proof.** The incrementality is immediate since given any two defeasible theories $D_1$ and $D_2$, $elim\_dft(D_1 \cup D_2) = elim\_dft(D_1) \cup elim\_dft(D_2)$.

To show that it is not modular, let us consider $D_1 = \{\leadsto p\}$, $D_2 = \{\Rightarrow \neg p\}$ and $\Sigma = \{p\}$. It is immediate to see that $D_1 \cup D_2 \vdash -\partial \neg p$. $elim\_dft(D_1) = \{\Rightarrow \neg p^-\}$, then $elim\_dft(D_1) \cup D_2 \vdash +\partial \neg p$. Therefore $D_1 \cup D_2 \not\equiv_\Sigma elim\_dft(D_1) \cup D_2$. $\square$

The next result shows that, in fact, we cannot eliminate defeaters in a modular way. In the following a set of rules $R$ denotes the defeasible theory $(\emptyset, R, \emptyset)$ (which may be obtained by application of *normal* followed by *elim_sup*).

**Theorem 17** *There is no modular transformation that transforms every defeasible theory $D$ into a theory $D'$ such that there are no defeaters in $D'$.*

**Proof.**

The claim of the theorem follows directly from the following auxiliary claim.

*Auxiliary Lemma:* Let $A \leadsto p$ be a defeater. Then, in general, there is no defeasible theory $R'$ without defeaters, such that for all defeasible theories $R$, $R \cup \{A \leadsto p\}$ and $R \cup R'$ allow the same conclusions in the language $\Sigma$ (where $\Sigma$ is the language of $R \cup \{A \leadsto p\}$).

*Proof of the Auxiliary Lemma:* Suppose there was such an $R'$ for the defeater $\leadsto p$. We will consider three different sets $R$:

1. $R = \emptyset$. Since $R'$ behaves the same as $\{\leadsto p\}$ we have:

    $R' \vdash -\partial p$, and

    $R' \vdash -\partial \neg p$.

2. $R = \{\Rightarrow p\}$. Since $R' \cup \{\Rightarrow p\}$ behaves the same as $\{\leadsto p, \Rightarrow p\}$ we have:

    $R' \cup \{\Rightarrow p\} \vdash +\partial p$, and

    $R' \cup \{\Rightarrow p\} \vdash -\partial \neg p$.

3. $R = \{\Rightarrow \neg p\}$. Since $R' \cup \{\Rightarrow \neg p\}$ behaves the same as $\{\leadsto p, \Rightarrow \neg p\}$ we have:

    $R' \cup \{\Rightarrow \neg p\} \vdash -\partial p$, and

    $R' \cup \{\Rightarrow \neg p\} \vdash -\partial \neg p$.

Let us first consider $R' \cup \{\Rightarrow p\}$. Consider a proof $P$ in $R' \cup \{\Rightarrow p\}$ of length $i+1$, such that $+\partial p$ is its last line, and $+\partial p$ does not occur in $P(1..i)$. By condition (2.3) in the definition of a proof[4], for every rule $r$ with consequent $\neg p$ there is a $b \in A(r)$ such that $-\partial b \in P(1..i)$.

Now we ask the following question: Can we regard $P(1..i)$ as a proof in $R'$? The only difference is that now the rule $\Rightarrow p$ is missing. What is the contribution of this rule in $P(1..i)$? Inspection of the definition of a proof shows that the rule is only used to add a line containing either $p$ or $\neg p$. In our particular case, given that only $+\partial p$ and $-\partial \neg p$ are derivable[5] and given that $+\partial p$ doesn't appear in $P(1..i)$, the only possible contribution of the rule $\Rightarrow p$ is to derive $-\partial \neg p$ somewhere in $P(1..i)$. Now we proceed as follows:

Case 1: $-\partial \neg p$ doesn't occur in $P(1..i)$. Then it can be shown by a simple induction on the length of $P$ that $P' = P$ is also a proof in $R'$.

Case 2: $-\partial \neg p$ occurs in $P(1..i)$. Then define $P'$ as follows: We know that $-\partial \neg p$ is derivable in $R'$. Take such a proof $P''$. Concatenate $P''$ and $P$ to construct $P'$[6]. Again it can be easily proven by induction on the length of proof that $P'$ is a proof in $R'$. Intuitively what we did was the following: The missing rule $\Rightarrow p$ may only cause problems in deriving $-\partial \neg p$ in $P$. But already we know that $-\partial \neg p$ is derivable in $R'$, so we establish this conclusion first and then proceed as in $P(1..i)$.

In both cases we get a proof $P'$ in $R'$ with the following property:

(∗) For every rule $r \in R'[\neg p]$ there is a $b \in A(r)$ such that $R' \vdash -\partial b$.

Now we turn our attention to $R' \cup \{\Rightarrow \neg p\}$. Despite the presence of $\Rightarrow \neg p$ which has no antecedents, $-\partial \neg p$ is derivable. Let $P$ be a proof of length $i+1$ with last line $-\partial \neg p$, such that $-\partial \neg p$ does not occur in $P(1..i)$. By the definition of a proof, there exists a rule $r$ in $R'[p]$ such that for all $a \in A(r)$ $+\partial a \in P(1..i)$.

Using the same argument as before[7], we can transform $P(1..i)$ to a proof $P'$ in $R'$, such that there exists a rule $s$ in $R'[p]$ such that for all $a \in A(s)$ $+\partial a \in P'$. Thus we have:

(∗∗) There exists a rule $s$ in $R'[p]$ such that for all $a \in A(s)$, $R' \vdash +\partial a$.

Obviously $R' \vdash -\Delta \neg p$ because $\{\leadsto p\} \vdash -\Delta \neg p$. Properties (∗) and (∗∗), together with the condition $+\partial$ in the definition of a proof, show that $R' \vdash +\partial p$. But also $R' \vdash -\partial p$ because $\{\leadsto p\} \vdash -\partial p$. [6] has shown that it is impossible to derive both together. Thus we have a contradiction. □

## 5.4 A Minimal Set of Ingredients

We have seen transformations that: (i) normalize; (ii) eliminate defeaters; and (iii) empty the superiority relation. First we summarize the outcome of our considerations.

**Theorem 18** *For every well-formed defeasible theory $D = (F, R, >)$ in the language $\Sigma$ we can effectively construct a normalized defeasible theory $D' = (\emptyset, R', \emptyset)$, such that $D$ and $D'$ have the same conclusions in $\Sigma$ and $R'_{dft} = \emptyset$.*

---

[4]Note that $\{\leadsto p, \Rightarrow p\} \nvdash +\Delta p$, thus $+\Delta p \notin P(1..i)$.

[5]As [6] shows, it is impossible to derive both $+\partial p$ and $-\partial p$.

[6]As [6] shows, by concatenating two proof of a defeasible theory $D$ one gets another proof in $D$.

[7]Essentially we are faced with the same situation: $P(1..i)$ is a proof in $R' \cup \{\Rightarrow \neg p\}$, we remove the rule $\Rightarrow \neg p$, and $-\partial \neg p$ doesn't occur in $P(1..i)$. So the only possible contribution of $\Rightarrow \neg p$ is to help derive $-\partial p$. But $-\partial p$ is already derivable in $R'$.

**Proof.** The effective procedure that transforms $D$ to $D'$ is the successive application of the three transformations we have already described: 1. *normal* 2. *elim_dft* 3. *elim_sup*. □

As a result of our discussion we may view a defeasible theory as a set $R$ of strict and defeasible rules. Next we show that none of the remaining ingredients used in a defeasible theory, strict rules and defeasible rules, can be eliminated while maintaining the set of conclusions. This should not come as a surprise, of course, since there is a technical as well as motivational/philosophical distinction between provability based on certain, definite knowledge only, and defeasible, nonmonotonic provability based on plausible assumptions, represented as defeasible rules.

**Proposition 19** *There is no correct transformation that eliminates strict rules.*

**Proof.** Suppose that there was such a correct transformation $T$. Consider the defeasible theory $R$ that consists only of the strict rule $\to p$. We have $\{\to p\} \vdash +\Delta p$. But since there is no strict rule in $T(R)$, $T(R) \nvdash +\Delta p$, which gives us a contradiction to the statement of the Proposition. □

**Proposition 20** *There is no correct transformation that eliminates defeasible rules.*

**Proof.** Suppose there was such a correct transformation $T$. Consider the defeasible theory $R$ that consists only of the defeasible rule $\Rightarrow p$. Then $\{\Rightarrow p\} \vdash +\partial p$. According to the claim of the Proposition $T(R) \vdash +\partial p$. But $T(R)$ consists only of strict rules. Inspection of the definition of the inference conditions in subsection 2.3 (a conclusion must be derived by a strict or a defeasible rule) shows that then also $T(R) \vdash +\Delta p$. But $\{\Rightarrow p\} \nvdash +\Delta p$, so we have a contradiction to the correctness of $T$. □

In the introduction we claimed that our results are useful for theoretical considerations, in addition to the implementational issues. The inference conditions in subsection 2.3 were rather complicated. In the following we show how the inference conditions $+\partial$ and $-\partial$ are simplified after the transformations have been applied. The reduced complexity is beneficial both for understanding the logic and for proofs.

$+\partial$: If $P(i+1) = +\partial q$ then either
    (1) $+\Delta q \in P(1..i)$ or
    (2) (2.1) $\exists r \in R[q] \forall a \in A(r) : +\partial a \in P(1..i)$ and
        (2.2) $-\Delta{\sim}q \in P(1..i)$ and
        (2.3) $\forall s \in R[{\sim}q] \exists a \in A(s) : -\partial a \in P(1..i)$

$-\partial$: If $P(i+1) = -\partial q$ then
    (1) $-\Delta q \in P(1..i)$ and
    (2) (2.1) $\forall r \in R[q] \exists a \in A(r) : -\partial a \in P(1..i)$ or
        (2.2) $+\Delta{\sim}q \in P(1..i)$ or
        (2.3) $\exists s \in R[{\sim}q] \forall a \in A(s) : +\partial a \in P(1..i)$

# 6  Conclusion

Defeasible Logic is a sceptical nonmonotonic logic based on the use of logical rules and priorities between them. Its features provide for a very natural expression of many of the standard examples used to motivate other nonmonotonic logics [16]. Moreover recent work in several application domains has demonstrated that defeasible reasoning shows great promise to be useful in practice [11, 3, 18].

This paper studied transformations of defeasible theories. The main results showed how facts, defeaters and the superiority relation can be simulated by the other ingredients of the logic. In doing so our focus was on transformations that satisfy modularity and incrementality conditions. The reason is that we should not think of a theory as a stand-alone representation of knowledge, but rather as a module to which rules can be added (or deleted). In such cases it is desirable for changes to be made on a bit-by-bit basis, and to be able to modify a part of a theory independently from the remainder.

One main consequence of these results is that we can study, without loss of generality, a simpler form of Defeasible Logic. Deeper results on a semantics for Defeasible Logic and the relationship between Defeasible Logic and other nonmonotonic and logic programming formalisms now become more accessible.

The other major benefit is in the implementation of systems. In fact our main transformations are utilized in an implementation of Defeasible Logic that has just been completed [13]. The implementation relies on a linear time algorithm to compute all conclusions from a defeasible theory without defeaters, and with an empty superiority relation. An input theory is transformed into this normal form by applying the transformations of section 5. It is worth noting that the transformations cause only a linear increase in the size of the defeasible theory (to be more precise, by a factor of 3 for normalization, a factor of 4 for the superiority relation, and a factor of 3 for the defeaters). As a result, we have a system that computes all conclusions in time linear in the size of the defeasible theory.

# Acknowledgments

This paper extends and revises work presented at the 11th Australian Joint Conference on Artificial Intelligence, and the 1998 Joint International Conference and Symposium on Logic Programming. This research was supported by the Australia Research Council under Large Grant No. A49803544.

# References

[1] G. Antoniou. *Nonmonotonic Reasoning*. MIT Press 1997.

[2] G. Antoniou, D. Billington and M.J. Maher. Normal Forms for Defeasible Logic. In *Proc. 1998 Joint International Conference and Symposium on Logic Programming*, MIT Press 1998, 160–174.

[3] G. Antoniou, D. Billington and M.J. Maher. On the analysis of regulations using defeasible rules. In *Proc. 32nd Hawaii International Conference on Systems Science*, 1999.

[4] G. Antoniou, F. Maruyama, R. Masuoka and H. Kitajima. Issues in Intelligent Information Integration. In *Proc. 3rd IASTED International Conference on Internet and Multimedia Systems and Applications*, IASTED 1999, 345-349.

[5] G. Antoniou, M.J. Maher and D. Billington. Defeasible Logic versus Logic Programming without Negation as Failure. *Journal of Logic Programming* 41,1 (2000): 45–57.

[6] D. Billington. Defeasible Logic is Stable. *Journal of Logic and Computation* 3 (1993): 370–400.

[7] E.F. Codd. Further Normalization of the Data Base Relational Model. In *Data Base Systems, Courant Computer Science Symposia Series 6*, Prentice Hall 1972.

[8] M.A. Covington, D. Nute and A. Vellino. *Prolog Programming in Depth*. Prentice Hall 1997.

[9] Y. Dimopoulos and A. Kakas. Logic Programming without Negation as Failure. In *Proc. 5th International Symposium on Logic Programming*, MIT Press 1995, 369–384.

[10] B.N. Grosof. Prioritized Conflict Handling for Logic Programs. In *Proc. Int. Logic Programming Symposium*, J. Maluszynski (Ed.), 197–211. MIT Press, 1997.

[11] B.N. Grosof, Y. Labrou and H.Y. Chan. A Declarative Approach to Business Rules in Contracts: Courteous Logic Programs in XML. In *Proc. 1st ACM Conference on Electronic Commerce (EC-99)*, ACM Press 1999.

[12] R.A. Kowalski and F. Toni. Abstract Argumentation, *Artificial Intelligence and Law Journal* 4(3-4), Kluwer Academic Publishers 1996.

[13] M.J. Maher, A. Rock, G. Antoniou, D. Billington and T. Miller. Efficient defeasible reasoning systems. Submitted to the *1st International Conference on Computational Logic*, 2000.

[14] V. Marek and M. Truszczynski. *Nonmonotonic Reasoning*. Springer 1993.

[15] D. Nute. Defeasible Reasoning. In *Proc. 20th Hawaii International Conference on Systems Science*, IEEE Press 1987, 470–477.

[16] D. Nute. Defeasible Logic. In D.M. Gabbay, C.J. Hogger and J.A. Robinson (Eds.): *Handbook of Logic in Artificial Intelligence and Logic Programming Vol. 3*, Oxford University Press 1994, 353–395.

[17] A. Pettorossi and M. Proietti. Transformation of Logic Programs: Foundations and Techniques. *Journal of Logic Programming* 19/20, 1994, 261–320.

[18] D.M. Reeves, B.N. Grosof, M.P. Wellman, and H.Y. Chan. Towards a Declarative Language for Negotiating Executable Contracts, *Proceedings of the AAAI-99 Workshop on Artificial Intelligence in Electronic Commerce (AIEC-99)*, AAAI Press / MIT Press, 1999.

[19] J.A. Robinson. A machine oriented logic based on the resolution principle. *Journal of the ACM* 12,1, 1965, 23–41.

[20] A. Rock. *Deimos: Query Answering Defeasible Logic System*. http://www.cit.gu.edu.au/~arock/defeasible/Defeasible.cgi
30